\documentclass[
aps,prd,twocolumn,superscriptaddress,
nofootinbib,amsmath,amssymb
]{revtex4-2}

\usepackage{orcidlink}
\usepackage{bm}
\usepackage{dcolumn}
\usepackage{graphicx}

\usepackage{adsmacros}

\newcommand{\ddif}{\mathrm{d}}
\newcommand{\lya}{Ly$\alpha$}
\newcommand{\lyaf}{Ly$\alpha$ forest}

\newcommand{\lyaxlya}{\lya$\times$\lya}
\newcommand{\lyaxqso}{\lya$\times\mathrm{QSO}$}

\newcommand{\hmpc}{$h^{-1}~$Mpc}

\newcommand{\aiso}{\alpha_\mathrm{iso}}
\newcommand{\aAP}{\alpha_\mathrm{AP}}

\defcitealias{desiY3LyaBAO2025}{DESI-DR2-Ly$\alpha$}
\defcitealias{collaborationPlanck2018Results2020}{Planck2018}

\begin{document}
\title{DESI DR2 Baryon Acoustic Oscillations from the Lyman Alpha Forest Multipoles}

\author{Naim~G\"oksel~Kara\c{c}ayl{\i}\orcidlink{0000-0001-7336-8912}}
\email{karacayli.1@osu.edu}
\affiliation{Center for Cosmology and AstroParticle Physics, The Ohio State University, 191 West Woodruff Avenue, Columbus, OH 43210, USA}
\affiliation{Department of Astronomy, The Ohio State University, 4055 McPherson Laboratory, 140 W 18th Avenue, Columbus, OH 43210, USA}
\affiliation{Department of Physics, The Ohio State University, 191 West Woodruff Avenue, Columbus, OH 43210, USA}

\author{Andrei~Cuceu\orcidlink{0000-0002-2169-0595}}
\affiliation{Lawrence Berkeley National Laboratory, 1 Cyclotron Road, Berkeley, CA 94720, USA}

\author{J.~Aguilar}
\affiliation{Lawrence Berkeley National Laboratory, 1 Cyclotron Road, Berkeley, CA 94720, USA}

\author{S.~Ahlen\orcidlink{0000-0001-6098-7247}}
\affiliation{Department of Physics, Boston University, 590 Commonwealth Avenue, Boston, MA 02215 USA}

\author{S.~Bailey\orcidlink{0000-0003-4162-6619}}
\affiliation{Lawrence Berkeley National Laboratory, 1 Cyclotron Road, Berkeley, CA 94720, USA}

\author{S.~BenZvi\orcidlink{0000-0001-5537-4710}}
\affiliation{Department of Physics \& Astronomy, University of Rochester, 206 Bausch and Lomb Hall, P.O. Box 270171, Rochester, NY 14627-0171, USA}

\author{D.~Bianchi\orcidlink{0000-0001-9712-0006}}
\affiliation{Dipartimento di Fisica ``Aldo Pontremoli'', Universit\`a degli Studi di Milano, Via Celoria 16, I-20133 Milano, Italy}
\affiliation{INAF-Osservatorio Astronomico di Brera, Via Brera 28, 20122 Milano, Italy}

\author{A.~Brodzeller\orcidlink{0000-0002-8934-0954}}
\affiliation{Lawrence Berkeley National Laboratory, 1 Cyclotron Road, Berkeley, CA 94720, USA}

\author{D.~Brooks}
\affiliation{Department of Physics \& Astronomy, University College London, Gower Street, London, WC1E 6BT, UK}

\author{T.~Claybaugh}
\affiliation{Lawrence Berkeley National Laboratory, 1 Cyclotron Road, Berkeley, CA 94720, USA}

\author{A.~de la Macorra\orcidlink{0000-0002-1769-1640}}
\affiliation{Instituto de F\'{\i}sica, Universidad Nacional Aut\'{o}noma de M\'{e}xico,  Circuito de la Investigaci\'{o}n Cient\'{\i}fica, Ciudad Universitaria, Cd. de M\'{e}xico  C.~P.~04510,  M\'{e}xico}

\author{B.~Dey\orcidlink{0000-0002-5665-7912}}
\affiliation{Department of Astronomy \& Astrophysics, University of Toronto, Toronto, ON M5S 3H4, Canada}
\affiliation{Department of Physics \& Astronomy and Pittsburgh Particle Physics, Astrophysics, and Cosmology Center (PITT PACC), University of Pittsburgh, 3941 O'Hara Street, Pittsburgh, PA 15260, USA}

\author{P.~Doel}
\affiliation{Department of Physics \& Astronomy, University College London, Gower Street, London, WC1E 6BT, UK}

\author{J.~Estrada}
\affiliation{Fermi National Accelerator Laboratory, PO Box 500, Batavia, IL 60510, USA}

\author{S.~Ferraro\orcidlink{0000-0003-4992-7854}}
\affiliation{Lawrence Berkeley National Laboratory, 1 Cyclotron Road, Berkeley, CA 94720, USA}
\affiliation{University of California, Berkeley, 110 Sproul Hall \#5800 Berkeley, CA 94720, USA}

\author{A.~Font-Ribera\orcidlink{0000-0002-3033-7312}}
\affiliation{Instituci\'{o} Catalana de Recerca i Estudis Avan\c{c}ats, Passeig de Llu\'{\i}s Companys, 23, 08010 Barcelona, Spain}
\affiliation{Institut de F\'{i}sica d’Altes Energies (IFAE), The Barcelona Institute of Science and Technology, Edifici Cn, Campus UAB, 08193, Bellaterra (Barcelona), Spain}

\author{J.~E.~Forero-Romero\orcidlink{0000-0002-2890-3725}}
\affiliation{Departamento de F\'isica, Universidad de los Andes, Cra. 1 No. 18A-10, Edificio Ip, CP 111711, Bogot\'a, Colombia}
\affiliation{Observatorio Astron\'omico, Universidad de los Andes, Cra. 1 No. 18A-10, Edificio H, CP 111711 Bogot\'a, Colombia}

\author{E.~Gaztañaga\orcidlink{0000-0001-9632-0815}}
\affiliation{Institut d'Estudis Espacials de Catalunya (IEEC), c/ Esteve Terradas 1, Edifici RDIT, Campus PMT-UPC, 08860 Castelldefels, Spain}
\affiliation{Institute of Cosmology and Gravitation, University of Portsmouth, Dennis Sciama Building, Portsmouth, PO1 3FX, UK}
\affiliation{Institute of Space Sciences, ICE-CSIC, Campus UAB, Carrer de Can Magrans s/n, 08913 Bellaterra, Barcelona, Spain}

\author{S.~Gontcho A Gontcho\orcidlink{0000-0003-3142-233X}}
\affiliation{University of Virginia, Department of Astronomy, Charlottesville, VA 22904, USA}

\author{A.~X.~Gonzalez-Morales\orcidlink{0000-0003-4089-6924}}
\affiliation{Departamento de F\'{\i}sica, DCI-Campus Le\'{o}n, Universidad de Guanajuato, Loma del Bosque 103, Le\'{o}n, Guanajuato C.~P.~37150, M\'{e}xico}

\author{G.~Gutierrez}
\affiliation{Fermi National Accelerator Laboratory, PO Box 500, Batavia, IL 60510, USA}

\author{C.~Hahn\orcidlink{0000-0003-1197-0902}}
\affiliation{Department of Astronomy, The University of Texas at Austin, 2515 Speedway Boulevard, Austin, TX 78712, USA}

\author{H.~K.~Herrera-Alcantar\orcidlink{0000-0002-9136-9609}}
\affiliation{Institut d'Astrophysique de Paris. 98 bis boulevard Arago. 75014 Paris, France}
\affiliation{IRFU, CEA, Universit\'{e} Paris-Saclay, F-91191 Gif-sur-Yvette, France}

\author{K.~Honscheid\orcidlink{0000-0002-6550-2023}}
\affiliation{Center for Cosmology and AstroParticle Physics, The Ohio State University, 191 West Woodruff Avenue, Columbus, OH 43210, USA}
\affiliation{Department of Physics, The Ohio State University, 191 West Woodruff Avenue, Columbus, OH 43210, USA}

\author{C.~Howlett\orcidlink{0000-0002-1081-9410}}
\affiliation{School of Mathematics and Physics, University of Queensland, Brisbane, QLD 4072, Australia}

\author{M.~Ishak\orcidlink{0000-0002-6024-466X}}
\affiliation{Department of Physics, The University of Texas at Dallas, 800 W. Campbell Rd., Richardson, TX 75080, USA}

\author{R.~Joyce\orcidlink{0000-0003-0201-5241}}
\affiliation{NSF NOIRLab, 950 N. Cherry Ave., Tucson, AZ 85719, USA}

\author{R.~Kehoe}
\affiliation{Department of Physics, Southern Methodist University, 3215 Daniel Avenue, Dallas, TX 75275, USA}

\author{D.~Kirkby\orcidlink{0000-0002-8828-5463}}
\affiliation{Department of Physics and Astronomy, University of California, Irvine, 92697, USA}

\author{T.~Kisner\orcidlink{0000-0003-3510-7134}}
\affiliation{Lawrence Berkeley National Laboratory, 1 Cyclotron Road, Berkeley, CA 94720, USA}

\author{A.~Kremin\orcidlink{0000-0001-6356-7424}}
\affiliation{Lawrence Berkeley National Laboratory, 1 Cyclotron Road, Berkeley, CA 94720, USA}

\author{O.~Lahav\orcidlink{0000-0002-1134-9035}}
\affiliation{Department of Physics \& Astronomy, University College London, Gower Street, London, WC1E 6BT, UK}

\author{M.~Landriau\orcidlink{0000-0003-1838-8528}}
\affiliation{Lawrence Berkeley National Laboratory, 1 Cyclotron Road, Berkeley, CA 94720, USA}

\author{J.M.~Le~Goff}
\affiliation{IRFU, CEA, Universit\'{e} Paris-Saclay, F-91191 Gif-sur-Yvette, France}

\author{L.~Le~Guillou\orcidlink{0000-0001-7178-8868}}
\affiliation{Sorbonne Universit\'{e}, CNRS/IN2P3, Laboratoire de Physique Nucl\'{e}aire et de Hautes Energies (LPNHE), FR-75005 Paris, France}

\author{M.~E.~Levi\orcidlink{0000-0003-1887-1018}}
\affiliation{Lawrence Berkeley National Laboratory, 1 Cyclotron Road, Berkeley, CA 94720, USA}

\author{M.~Manera\orcidlink{0000-0003-4962-8934}}
\affiliation{Departament de F\'{i}sica, Serra H\'{u}nter, Universitat Aut\`{o}noma de Barcelona, 08193 Bellaterra (Barcelona), Spain}
\affiliation{Institut de F\'{i}sica d’Altes Energies (IFAE), The Barcelona Institute of Science and Technology, Edifici Cn, Campus UAB, 08193, Bellaterra (Barcelona), Spain}

\author{P.~Martini\orcidlink{0000-0002-4279-4182}}
\affiliation{Center for Cosmology and AstroParticle Physics, The Ohio State University, 191 West Woodruff Avenue, Columbus, OH 43210, USA}
\affiliation{Department of Astronomy, The Ohio State University, 4055 McPherson Laboratory, 140 W 18th Avenue, Columbus, OH 43210, USA}

\author{A.~Meisner\orcidlink{0000-0002-1125-7384}}
\affiliation{NSF NOIRLab, 950 N. Cherry Ave., Tucson, AZ 85719, USA}

\author{R.~Miquel}
\affiliation{Instituci\'{o} Catalana de Recerca i Estudis Avan\c{c}ats, Passeig de Llu\'{\i}s Companys, 23, 08010 Barcelona, Spain}
\affiliation{Institut de F\'{i}sica d’Altes Energies (IFAE), The Barcelona Institute of Science and Technology, Edifici Cn, Campus UAB, 08193, Bellaterra (Barcelona), Spain}

\author{J.~Moustakas\orcidlink{0000-0002-2733-4559}}
\affiliation{Department of Physics and Astronomy, Siena University, 515 Loudon Road, Loudonville, NY 12211, USA}

\author{A.~Muñoz-Gutiérrez}
\affiliation{Instituto de F\'{\i}sica, Universidad Nacional Aut\'{o}noma de M\'{e}xico,  Circuito de la Investigaci\'{o}n Cient\'{\i}fica, Ciudad Universitaria, Cd. de M\'{e}xico  C.~P.~04510,  M\'{e}xico}

\author{S.~Nadathur\orcidlink{0000-0001-9070-3102}}
\affiliation{Institute of Cosmology and Gravitation, University of Portsmouth, Dennis Sciama Building, Portsmouth, PO1 3FX, UK}

\author{N.~Palanque-Delabrouille\orcidlink{0000-0003-3188-784X}}
\affiliation{IRFU, CEA, Universit\'{e} Paris-Saclay, F-91191 Gif-sur-Yvette, France}
\affiliation{Lawrence Berkeley National Laboratory, 1 Cyclotron Road, Berkeley, CA 94720, USA}

\author{W.~J.~Percival\orcidlink{0000-0002-0644-5727}}
\affiliation{Department of Physics and Astronomy, University of Waterloo, 200 University Ave W, Waterloo, ON N2L 3G1, Canada}
\affiliation{Perimeter Institute for Theoretical Physics, 31 Caroline St. North, Waterloo, ON N2L 2Y5, Canada}
\affiliation{Waterloo Centre for Astrophysics, University of Waterloo, 200 University Ave W, Waterloo, ON N2L 3G1, Canada}

\author{F.~Prada\orcidlink{0000-0001-7145-8674}}
\affiliation{Instituto de Astrof\'{i}sica de Andaluc\'{i}a (CSIC), Glorieta de la Astronom\'{i}a, s/n, E-18008 Granada, Spain}

\author{I.~P\'erez-R\`afols\orcidlink{0000-0001-6979-0125}}
\affiliation{Departament de F\'isica, EEBE, Universitat Polit\`ecnica de Catalunya, c/Eduard Maristany 10, 08930 Barcelona, Spain}

\author{G.~Rossi}
\affiliation{Department of Physics and Astronomy, Sejong University, 209 Neungdong-ro, Gwangjin-gu, Seoul 05006, Republic of Korea}

\author{E.~Sanchez\orcidlink{0000-0002-9646-8198}}
\affiliation{CIEMAT, Avenida Complutense 40, E-28040 Madrid, Spain}

\author{D.~Schlegel}
\affiliation{Lawrence Berkeley National Laboratory, 1 Cyclotron Road, Berkeley, CA 94720, USA}

\author{M.~Schubnell}
\affiliation{Department of Physics, University of Michigan, 450 Church Street, Ann Arbor, MI 48109, USA}
\affiliation{University of Michigan, 500 S. State Street, Ann Arbor, MI 48109, USA}

\author{H.~Seo\orcidlink{0000-0002-6588-3508}}
\affiliation{Department of Physics \& Astronomy, Ohio University, 139 University Terrace, Athens, OH 45701, USA}

\author{J.~Silber\orcidlink{0000-0002-3461-0320}}
\affiliation{Lawrence Berkeley National Laboratory, 1 Cyclotron Road, Berkeley, CA 94720, USA}

\author{D.~Sprayberry}
\affiliation{NSF NOIRLab, 950 N. Cherry Ave., Tucson, AZ 85719, USA}

\author{G.~Tarl\'{e}\orcidlink{0000-0003-1704-0781}}
\affiliation{University of Michigan, 500 S. State Street, Ann Arbor, MI 48109, USA}

\author{B.~A.~Weaver}
\affiliation{NSF NOIRLab, 950 N. Cherry Ave., Tucson, AZ 85719, USA}

\author{R.~Zhou\orcidlink{0000-0001-5381-4372}}
\affiliation{Lawrence Berkeley National Laboratory, 1 Cyclotron Road, Berkeley, CA 94720, USA}

\author{H.~Zou\orcidlink{0000-0002-6684-3997}}
\affiliation{National Astronomical Observatories, Chinese Academy of Sciences, A20 Datun Road, Chaoyang District, Beijing, 100101, P.~R.~China}

\date{\today}

\begin{abstract}
We present an alternative measurement of the Baryon Acoustic Oscillation (BAO) using the Legendre multipole representation of the Ly$\alpha$ forest correlation functions from the second data release (DR2) of the Dark Energy Spectroscopic Instrument survey. Compressing the auto- and cross-correlation functions into Legendre multipoles yields a positive-definite covariance matrix without any smoothing---unlike the baseline DR2 analysis---thanks to a significantly reduced data vector size. We introduce the statistical corrections required to debias the finite-sample covariance matrix estimate and demonstrate that monopole and quadrupole terms for both auto- and cross-correlations can be used even when the correlation functions are distorted by continuum errors and contaminated by metals. This formalism has slightly diminished the constraining power of the BAO scale, while considerably weakening constraints on nuisance parameters. We measure the isotropic BAO scale with $0.93\%$ precision at $z_\mathrm{eff}=2.35$, the Hubble parameter $H(z_\mathrm{eff})=(239.5\pm3.4)~(147.09~\mathrm{Mpc}/r_d) ~\mathrm{km~s}^{-1}~\text{Mpc}^{-1}$, and the transverse comoving distance $D_M(z_\mathrm{eff})=(5.80 \pm 0.10)~(r_d/147.09~\mathrm{Mpc})$~Gpc for a given value of the sound horizon ($r_d$). Our BAO results are entirely consistent with the baseline DR2 analysis.
\end{abstract}

\maketitle

\section{Introduction}
The Dark Energy Spectroscopic Instrument (DESI; \cite{leviDESIExperimentWhitepaper2013, desicollaborationDESIExperimentPart2016, desicollaborationDESIExperimentPart2016b, abareshiOverviewInstrumentationDark2022}) is currently delivering measurements of the baryon acoustic oscillation (BAO) scale with sub-percent precision \cite{desiKp4BaoGalaxies2024, desiKp6BaoLya2024, desiY3LyaBAO2025, desiY3BaoAndCosmology2025}. In combination with the cosmic microwave background (CMB) and type Ia supernovae (SNe), the Data Release 2 (DR2) results of DESI, with over 14 million galaxies and 1.2 million quasars, favor ``dynamical" dark energy models instead of the cosmological constant \cite{desiKp7Cosmology2024, desi2024VFullShape, desiY3BaoAndCosmology2025}. As DESI observations continue to improve, we revisit the choices for the Lyman-$\alpha$ (\lya) forest analysis, specifically the dimensionality of the correlation function covariance matrix, to ensure the robustness of current and future DESI results.

The \lyaf\ is the collection of the absorption lines of neutral hydrogen gas at wavelengths shorter than the \lya\ emission line. It provides a unique window into the matter distribution and the tightest constraints on the cosmic expansion rate at the redshift range $2<z<4$ \cite{desiKp6BaoLya2024, desiY3LyaBAO2025}. The BAO scale is extracted from the three-dimensional auto-correlation of \lyaf\ (\lyaxlya) and the cross-correlation between \lyaf\ and quasars (\lyaxqso) \cite{buscaBaryonAcousticOscillations2013, fontriberaQuasarLymanCrossBossDr112014, slosarMeasurementBaryonAcoustic2013, bourbouxCompletedSDSSIVExtended2020, desiKp6BaoLya2024, desiY3LyaBAO2025}. Beyond the BAO feature, the full shape of the correlation function carries additional cosmological information in the form of the Alcock-Paczynski (AP) effect \cite{alcock_paczynski1979} and redshift space distortions \cite{kaiser1987}. Including these leads to tighter constraints on the cosmic expansion rate and a measurement of the growth rate of cosmic structure, $f\sigma_8$ \cite{cuceuBeyondBao2021, cuceuCosmicExpansion2023, cuceuAP2023, cuceuDesiDr1FullShape2025}.

The standard approach measures the \lyaf\ correlation functions in bins of transverse and line-of-sight separations, $(r_\bot, r_\|)$. This basis is better suited to describe distortions due to quasar continuum-fitting errors and metal-line systematics, which predominantly affect small $r_\bot$ scales. However, it yields a larger data vector (2,500 elements for autocorrelations and 5,000 elements for cross-correlations) than the independent realizations possible with subsampling of real or mock data. This poses a significant challenge for covariance matrix estimation, since at least as many independent samples as the data vector size are needed to suppress random noise in the estimated covariance. In practice, an ad-hoc smoothing is applied to the noisy covariance estimates in $(r_\bot, r_\|)$, which is shown to be adequate for the current precision level of the data \cite{cuceuValidationDesiDr12025, casasValidationOfDesiDr2Lya2025}. 
However, this smoothing procedure is not guaranteed to work in the high signal-to-noise regime. For example, Fig.~11 of \citet{cuceuValidationDesiDr12025} visually indicates that the off-diagonal structure of the \lyaxqso\ covariance matrix changes with smoothing. Additionally, alternative continuum fitting methods \cite[e.g.,][]{turnerLyaForestMeanFluxFromDesiY12024}) in which the large-scale information is preserved rather than projected out may manifest a different structure than the one assumed in smoothing \cite{turnerLimitsOfCosmologicalInformation2025}.

To address these limitations, we explore an alternative measurement scheme that dramatically reduces the dimensionality of the correlation function data vector by decomposing it into Legendre multipoles. Multipole compression is a well-established technique in galaxy and quasar clustering analyses \cite{sdssClusteringDr122017, desiKp4BaoGalaxies2024, desiY3BaoAndCosmology2025} and has been previously considered in the context of \lya\ forest correlations \cite{buscaBaryonAcousticOscillations2013}. 
Nevertheless, it has been set aside due to its limitations in describing distortions arising from continuum fitting errors and metal contamination. For example, ionized silicon systems cause offsets in the correlation function most strongly along the line of sight (small $r_\bot$). In principle, a multipole representation that uses all components would work equally well. However, when the multipoles are truncated to $\ell \leq 4$, which is our main focus for reducing the data size, this effect is diluted by angular integration.
Despite these challenges, we argue that the reduced data dimensionality improves the covariance matrix estimation and makes the multipole-based analysis a valuable alternative. By connecting \lyaf\ analyses to the well-developed galaxy clustering literature, this approach has the potential to enhance the robustness of future DESI \lyaf\ results.


This paper is organized as follows. We provide an overview of the DR2 \lyaf\ data, including catalogs of astrophysical contaminants such as damped \lya\ absorption systems and broad absorption line quasars, in Sec.~\ref{sec:data}. Sec.~\ref{sec:method} describes our method for measuring the correlation function multipoles and the necessary changes to the existing pipeline. This section also includes a summary of the modeling of correlations. We validate our method, investigate the optimal number of multipoles, and the goodness of fit using ten mock realizations in Sec.~\ref{sec:validation}. We present our BAO measurements using this technique and compare them with the baseline DR2 results in Sec.~\ref{sec:results}. We discuss advantages, disadvantages, and possible extensions in Sec.~\ref{sec:discuss}.

\section{Data\label{sec:data}}
Our sample is the processed data from the first three years of the DESI survey, which will be publicly available in the future as DR2. DESI DR2 covers a larger area than DR1 \cite{desiKp2DataRelease12024} and has nearly 1.3 million quasar spectra with $z>1.77$ and 825,000 with $z>2.09$, typically observed 2--3 times. This dataset was used in the key science paper measuring the BAO scale from the \lyaf\ \cite[\citetalias{desiY3LyaBAO2025} from now on]{desiY3LyaBAO2025}. The instrumentation and connection to the science requirements are overviewed in ref.~\cite{abareshiOverviewInstrumentationDark2022}, with additional papers detailing the robotic focal plane assembly \cite{silberRoboticMultiobjectFocal2023}, the wide-field corrector system \cite{millerCorrector2024}, the highly-efficient fiber system \cite{poppettFiberSystem2024}, and the survey operations \cite{schlaflySurveyOps}. 
A sophisticated data processing pipeline helps to minimize instrumental systematic errors \cite{guySpectroscopicDataProcessingPipeline2022}. Multiple observations of a given quasar are coadded and organized by their \texttt{HEALPix} pixel \cite{healpix}. 
This pixelization scheme is also used to split correlation function measurements into subsamples when estimating the covariance matrix.

The \lyaf\ measurements are contaminated by several astrophysical effects. Among them, damped \lya\ absorption (DLA) systems and broad absorption lines (BALs) are identified and masked out from the analysis.
DLAs have neutral hydrogen column densities $N_\mathrm{HI} > 2 \times 10^{20}~\text{cm}^{-2}$ and are denser objects that are more strongly clustered than the \lyaf\ \cite{font-riberaLargeScaleCrossCorrelationsDlaForest2012}. They are also saturated absorption lines with damping wings extending to large separations, which impact the continuum level. DLAs are identified, and their column densities and redshifts are measured through a combination of a template-based approach \cite{brodzellerConstructionDlaDesiDr2}, a convolutional neural network \cite{wangDeepLearningDESIDLA2022}, and a Gaussian process \cite{mingfengDLAGP2021} method.

BAL quasars have absorption features that arise from quasar outflows and are unrelated to the matter distribution in the intergalactic medium. These are identified in the same manner as the DESI DR1 analysis \cite{desiKp6BaoLya2024}. Ref.~\cite{martiniDesiBalY12024} describes the method in detail and demonstrates that incompleteness in the BAL catalog has minimal impact on the BAO measurements.

We use the same forest measurements of \citetalias{desiY3LyaBAO2025} that are in the 3600--5772~\AA\ wavelength range. The astrophysical contaminants, such as DLAs and BALs, cosmic rays, and major sky emission lines, are masked before the quasar continuum is estimated. Forests that are shorter than 120~\AA\ are discarded to ensure a sufficient length in continuum fitting. The transmitted flux fluctuations $\delta_q(\lambda)$ for each quasar $q$ are given by
\begin{equation}
    \delta_q(\lambda) = \frac{f_q(\lambda)}{\overline{F}(\lambda)C_q(\lambda)} - 1,
\end{equation}
where $\overline{F}$ is the mean intergalactic medium (IGM) transmission and $C_q$ is the quasar continuum. The fitting procedure is described in detail by ref.~\cite{ramirezperezLyaCatalogDesiEdr2023} and performed in the forest region. This removes the underlying large-scale density field from each forest, which distorts the measured \lya\ clustering statistics through mixing of short- and long-range correlations.

We analyze only Region A, which is defined as the 1040--1205~\AA\ wavelength range of the quasar rest-frame, and exclude Region B (920--1020~\AA) from our analysis. This is to ensure the size of our data vector remains small and to ascertain that the covariance matrix can be reliably estimated from the sub-samples. Excluding the B region has a minor impact, which will be demonstrated in Sec.~\ref{sec:validation}, since the measurements in this region are lower in SNR and contaminated by higher-order Lyman transitions. Nevertheless, Region B can be reintroduced through more advanced techniques of covariance matrix estimation in future work.

\section{Method\label{sec:method}}
In order to measure the multipoles of the correlation function, we first make minor adjustments to our previous approach. In \citetalias{desiY3LyaBAO2025} (and earlier work \cite{bautistajuliane.MeasurementBaryonAcoustic2017, bourbouxCompletedSDSSIVExtended2020, desiKp6BaoLya2024}), we employed a rectangular grid in directions parallel ($r_\|$) and transverse ($r_\bot$) to the line-of-sight in clustering measurements of the \lyaf. For a pixel pair $(i,j)$ with angular separation $\theta_{ij}$, these are defined as
\begin{align}
    r_\| &= \left( D_C(z_i) - D_C(z_j)\right)\cos (\theta_{ij}/2) \\
    r_\bot &= \left( D_M(z_i) + D_M(z_j)\right)\sin (\theta_{ij}/2),
\end{align}
where $D_C$ is the comoving distance and $D_M$ is the transverse comoving distance, which are identical for a flat $\Lambda$CDM cosmology. We adopt the same fiducial cosmology as the \citetalias{desiY3LyaBAO2025}, which is based on ref.~\cite{collaborationPlanck2018Results2020}.

We first modify the grid to measure the anisotropic two-point correlation function $\xi(r, \mu)$ at a separation distance $r=\sqrt{r_\|^2 + r_\bot^2}$ and $\mu=r_\|/r$. The estimator is the same weighted average for each $(r, \mu)$ bin, where the effective $r_\mathrm{eff}$, $\mu_\mathrm{eff}$, and $z_\mathrm{eff}$ are calculated using the same weights (a smoothed model is evaluated at these effective points as discussed later).

We use 50 $r$-bins of 4~\hmpc\ up to 200~\hmpc\ and 100 $\mu$ bins in $[0,1]$ for the auto-correlations and 200 $\mu$ bins in $[-1,1]$ for cross-correlations. The distortion matrices are calculated using the same bin sizes up to 300~\hmpc\ to capture the contributions from larger separations.\footnote{Including separations $r_\|>300$~\hmpc\ may improve the cross-correlation accuracy (see ref.~\cite{buscaEffectsOfContinuumFitting2025}). We prefer following the \citetalias{desiY3LyaBAO2025} as closely as possible in this work.}$^,$\footnote{The window function can be more accurately propagated by estimating the correlation function in narrower $r$ bins as done in DESI's galaxy and quasar clustering analysis \cite{desiKp2DataRelease12024, desiKp4BaoGalaxies2024, desiY3BaoAndCosmology2025}. However, the distortion matrix presents additional complications and increases the computational time.} Then, we project $\xi(r, \mu)$ onto the Legendre polynomial basis:
\begin{equation}
    \label{eq:multipole}\xi_\ell(r) = \frac{2\ell + 1}{2}\int \ddif\mu ~\xi(r, \mu) \mathcal{L}_\ell(\mu),
\end{equation}
where $\mathcal{L}_\ell(\mu)$ is the Legendre polynomial. Following ref.~\cite{desiKp4BaoGalaxies2024}, this integral is computed as a finite sum over $\mu$-bins, weighted by the integral $\frac{2\ell + 1}{2}\int \ddif\mu \mathcal{L}_\ell(\mu)$ within each bin. This projection is performed in the \texttt{Vega} package\footnote{\url{https://github.com/andreicuceu/vega}}, which models the correlations and infers parameters. It follows that the distortion and covariance (without smoothing in our case) matrices, as well as the model, are calculated as before in the full $(r, \mu)$ grid, and then projected onto the Legendre basis in a coherent manner. These minimal changes seamlessly integrate multipoles into our existing pipeline.

Our approach differs from ref.~\cite{buscaBaryonAcousticOscillations2013} in which the multipoles are calculated through a fit. Although this approach may enhance the signal by incorporating weights into the multipole compression, it also compromises the orthonormality of the multipoles, complicates the window function, and hinders a precise connection to the model, since the parameters differ across fitting setups. These least-squares solutions may be preferred in the low-signal regime, but are not needed at the statistical power of DR2.

The multipole representation of the correlation function dramatically reduces the dimension of our data vector, which forms the basis of our covariance matrix study. We estimate the covariance matrix using the sample covariance estimator over \texttt{HEALPix} pixels as before:
\begin{equation}
    \label{eq:cov} C_{AB} = \frac{\sum_h w^h_Aw^h_B (\xi^h_A - \bar\xi_A) (\xi^h_B - \bar \xi_B)}{\sum_h w^h_A \sum_h w^h_B},
\end{equation}
where the indices $A$ and $B$ represent $(r,\mu)$ bins. This estimator is formally biased and needs a sample bias correction. However, as we show in Appendix~\ref{app:sample_bias}, this correction is insignificant ($\sim0.1\%$) and can be ignored for DR2. This covariance matrix is then compressed in both axes to obtain the covariance of multipoles: $\mathbf{C} \rightarrow \mathbf{P}_\ell\mathbf{C}\mathbf{P}_\ell^\mathrm{T}$, where $\mathbf{P}_\ell$ is the matrix representation of the finite-sum Legendre decomposition. We choose $\texttt{nside}=16$ such that the average size of a pixel (250~\hmpc\ at $z=2.3$) is larger than the largest separation used in the correlation function (180~\hmpc), resulting in $N_h = 1147$ ``independent" samples for DR2.

The $(r_\bot, r_\|)$ binning yields a correlation function with over 7000 elements, the covariance of which becomes noisy since it cannot be reliably estimated using only about 1000 samples. In the baseline DESI analyses, we smooth this noisy covariance estimate by averaging the non-diagonal elements of the correlation matrix $(C_{AB}/\sqrt{C_{AA}C_{BB}})$ of the same $|r_\|^A-r_\|^B|$ and $|r_\bot^A-r_\bot^B|$ differences. In parallel, our formalism in $(r, \mu)$ binning still yields a noisy covariance matrix. In contrast, however, the Legendre compression of this matrix $(\mathbf{P}_\ell\mathbf{C}\mathbf{P}_\ell^\mathrm{T})$ is a well-motivated ``averaging" operation and not an assertion about the correlation matrix's structure. This allows us to focus on the corrections to the covariance matrix below.

Using the sample mean and covariance estimates of a Gaussian-distributed dataset for parameter inference leads to a non-Gaussian likelihood function with heavier tails, similar to the Student's t-distribution \cite{sellentinHeavensEstimatedCovariance2016}. To approximate this widening in the distribution, we multiply our covariance matrix by the inverse Hartlap factor \cite{hartlapInverseCovariance2007} and keep our likelihood Gaussian following ref.~\cite{desi2024VFullShape}:
\begin{equation}
    \label{eq:hartlap}C_\mathrm{Hartlap} = \frac{N_h - 1}{N_h - N_b - 2},
\end{equation}
where $N_h$ is the number of \texttt{HEALPix} samples, $N_b$ is the number of bins, and $\mathbf{C} \rightarrow C_\mathrm{Hartlap} \mathbf{C} $. Furthermore, we propagate the uncertainty in the covariance matrix to the inferred parameters by multiplying the \emph{parameter} covariance matrix by the following correction factor:
\begin{equation}
    \label{eq:percival}C_\mathrm{Percival} = \frac{1 + C_1(N_b - N_\mathrm{params})}{1 + C_0 + C_1(N_\mathrm{params} - 1)},
\end{equation}
where $N_\mathrm{params}$ is the number of inferred parameters and
\begin{align}
    C_0 &= \frac{2}{(N_h - N_b - 1) (N_h - N_b - 4)} \\
    C_1 &= \frac{N_h - N_b - 2}{2} C_0
\end{align}
as introduced by ref.~\cite{percivalIncludingCovarianceErrors2014}. $C_\mathrm{Hartlap}$ and $C_\mathrm{Percival}$ factors range between $1.1-1.25$ in our work.

When $N_h < N_b$ as is the case in \citetalias{desiY3LyaBAO2025}, these corrections become negative, hence inapplicable. This is another manifestation of insufficient samples for covariance matrix estimation, alongside non-positive-definite outcomes. However, these corrections themselves are approximations of the underlying true likelihood, not exact quantifications of biases in covariance matrix estimates. Similarly, the smoothing procedure in \citetalias{desiY3LyaBAO2025} serves as an approximation for the true likelihood and covariance matrix, replacing these corrections. It has been shown that this procedure yields unbiased errors on the fitted parameters using hundreds of survey realizations \cite{casasValidationOfDesiDr2Lya2025, cuceuValidationDesiDr12025}. We are concerned that it is unclear when this approximation will break down and what to replace it with, and we find that the truncated multipole representation offers a promising avenue.


\subsection{Modeling the correlations}
In this section, we provide an overview of our model, which was applied to DESI DR1 and DR2 \lyaf\ analyses to extract the BAO scale \cite{desiKp6BaoLya2024, desiY3LyaBAO2025} as well as the full-shape information \cite{cuceuDesiDr1FullShape2025}. The \lyaf\ correlation model is fundamentally based on the linear theory, where non-linearities are modeled with an empirical model built from simulations. The major modeling difficulties arise from distortions due to continuum fitting and contaminations from metal and high-column-density systems. As to our multipole representation, the full model is first computed on an $(r, \mu)$ grid, as in previous DESI analyses, and then projected onto the Legendre multipole basis, as mentioned in the previous section. This ensures that the model and data are consistently handled and the window function is accurately propagated. This model is publicly accessible in the \texttt{Vega} software package.

We calculate a template isotropic linear matter power spectrum using \texttt{CAMB} \cite{lewisCAMB2000} at the fiducial cosmology \cite{collaborationPlanck2018Results2020}. The BAO information is isolated by decomposing this into peak and smooth components, while applying a Gaussian smoothing to the peak component to model the non-linear broadening of the BAO feature as described by ref.~\cite{kirkbyFittingMethodsBao2013}:
\begin{equation}
    \label{eq:bao_broadening}P_\mathrm{QL}(k_\bot, k_\|) = P_\mathrm{peak}(k) e^{-(\Sigma_\bot^2 k_\bot^2 + \Sigma_\|^2 k_\|^2) / 2} + P_\mathrm{sm}(k).
\end{equation}
We introduce two BAO scaling parameters $(\alpha_\bot, \alpha_\|)$ that multiply the coordinates in transverse and parallel directions only for the peak component of the correlation function. As a deviation from previous DESI analyses, we directly measure the isotropic scaling parameter $\aiso=(\alpha_\|\alpha_\bot^2)^{1/3}$ and the Alcock-Paczynski parameter $\aAP=\alpha_\| / \alpha_\bot$, since $\aiso$ is the natural parameter constrained by the monopole. The $\alpha_\bot$ and $\alpha_\|$ parameters are derived from $\aiso$ and $\aAP$. 

For two tracers $X$ and $Y$, the anisotropic power spectrum is built on the Kaiser kernel at linear order $K_i(\mu) = b_i (1+\beta_i\mu^2)$ \cite{kaiser1987}, where $\mu \equiv k_\| / k$:
\begin{equation}
    \label{eq:model_pxy} P_{XY} = K_X(\mu) K_Y(\mu) P_\mathrm{QL}(k, \mu)  F_\mathrm{NL}^{XY}(k, \mu) G(k, \mu),
\end{equation}
where $F_\mathrm{NL}^{XY}$ is the aforementioned non-linear fitting fuction. There are two possibilities for $XY$ combinations in our work. For \lyaxlya\ auto-correlations, we use the fitting function in ref.~\cite{arinyoNonLinearPowerLya2015} for $F_\mathrm{NL}^{\alpha\alpha}$. For \lyaxqso\ cross correlations, we model the impact of quasar redshift errors and fingers-of-god effect due to peculiar velocities using a Lorentzian ($F_\mathrm{NL}^{Q\alpha}=\sqrt{1 + (k_\|\sigma_v)^2}$) for real data following \citetalias{desiY3LyaBAO2025}, but using a Gaussian ($F_\mathrm{NL}^{Q\alpha}=\exp\left[-(k_\|\sigma_v)^2/4\right]$) for mock data following ref.~\cite{casasValidationOfDesiDr2Lya2025}. When the tracer refers to quasars, the RSD parameter is given by $\beta_Q=f/b_Q$, where $f$ is the linear growth rate. Following previous \lyaf\ BAO analyses \cite{desiY3LyaBAO2025}, we fix $f$ to its fiducial value and marginalize over $b_Q$.\footnote{Previous analyses have found this does not have a significant impact on BAO measurements \cite{desiKp6BaoLya2024}, and measuring this parameter is generally only done in full-shape analyses \cite[e.g.,][]{cuceuDesiDr1FullShape2025}}. On the other hand, the \lya\ RSD parameter has an extra unknown velocity divergence bias, $b_\eta$, and is given by $\beta_\alpha=b_\eta f / b_\alpha$ \cite[see e.g.,][]{seljakKaiser,chenKaiser}. Therefore, in this case, we marginalize over both $b_\alpha$ and $\beta_\alpha$, following \cite{desiY3LyaBAO2025}.

The smoothing due to binning the correlation function is incorporated through the $G(k)$ term. We include the smoothing of radial bins by multiplying the model spectrum by $G(k)=\mathrm{sinc}(k\Delta r/2)$, which is a sufficient approximation for our investigation, as validation tests prove. The smoothing in $\mu$ bins is already included in the finite sum of the Legendre decomposition. We obtain the correlation function by transforming the first four even multipoles of the anisotropic power spectrum in Eq.~(\ref{eq:model_pxy}) using the FFTLog algorithm \cite{hamiltonUncorrelatedModesNonlinear2000}. The correlation function on the $(r, \mu)$ grid is then reconstructed from these multipoles, which will be reprojected onto the Legendre multipole basis after accounting for systematics and distortions.

Major model complications associated with the multipole representation arise from (1) residual HCDs, (2) metal contamination, and (3) distortions resulting from biased continuum estimates. HCDs are tracers of overdense regions of the matter field with a different bias, $b_\mathrm{HCD}$, than the forest. They exhibit saturated absorption lines with long damping wings, leading to a suppression in power along the line of sight. The linear order model follows a simple rescaling of $b_\alpha$ and $\beta_\alpha$:
\begin{align}
    b_\alpha &\rightarrow b_\alpha + b_\mathrm{HCD} F_\mathrm{HCD}(k_\|), \\ b_\alpha\beta_\alpha &\rightarrow b_\alpha\beta_\alpha + b_\mathrm{HCD}\beta_\mathrm{HCD} F_\mathrm{HCD}(k_\|),
\end{align}
where we approximate the suppression in the line of sight with $F_\mathrm{HCD}(k_\|) = \exp(-L_\mathrm{HCD} k_\|)$, which is the Fourier transform of the Lorentzian broadening in the Voight profile, and $L_\mathrm{HCD}\sim 5$~\hmpc\ is the typical length of the HCD absorption  \cite{font-riberaEffectHighColumn2012, rogersCorrelations3dHighColumnDensity2018, bourbouxCompletedSDSSIVExtended2020, tanModelingHcdSystems2025}. We note that the BAO broadening in eq.~(\ref{eq:bao_broadening}) is also applied to HCDs. In configuration space, the $F_\mathrm{HCD}(k_\|)$ term smooths the 2D correlation function, both the peak and smooth components, along the line-of-sight direction with a scale set by $L_\mathrm{HCD}$. In order to detect this change in shape, one needs high enough precision and/or sufficiently dense sampling of the correlation function. Consequently, our typical $\Delta r =4~$\hmpc\ bin size effectively sets a lower threshold for detecting HCDs even for full 2D analyses. Furthermore, we expect our multipole representation to be significantly less effective at constraining these HCD parameters. In fact, we fail to detect $b_\mathrm{HCD}$ in the real data application as discussed in Sec.~\ref{sec:results}.

In terms of the metal contamination, we follow the recipe in refs.~\cite{desiKp6BaoLya2024, desiY3LyaBAO2025}. Metal systems are misplaced in the radial direction because the \lya\ transition wavelength is used as opposed to their true transition wavelength. This results in an offset in the $r_\|$ direction of the correlation function. Our recipe loops over all possible pairs and computes the relative weight of pairs with true separations in bin $B$ among those with misestimated separations in bin $A$. These weights construct a metal matrix, $M_{AB}$, in the $(r_\bot, r_\|)$ grid that maps ``true" correlation function to a measured correlation, $\xi^\mathrm{meas}_A=\sum_B M_{AB} \xi^\mathrm{true}_B$, for each pair of transitions. We then interpolate $\xi^\mathrm{meas}$ from the $(r_\bot, r_\|)$ grid onto the $(r, \mu)$ grid.

The continuum fitting procedure introduces distortions to measured correlation functions \cite{slosarLya3DBOSS2011, bautistajuliane.MeasurementBaryonAcoustic2017, buscaEffectsOfContinuumFitting2025}. These are modeled with a distortion matrix, $\mathbf{D}^{XY}$, that multiplies the correlation function: $\bm{\xi}_\mathrm{dist}^{XY} = \mathbf{D}^{XY} \cdot \bm{\xi}_\mathrm{undist}^{XY}$. This distortion matrix varies for each $XY$ combination and is computed from the dataset on the $(r, \mu)$ grid, reflecting the weights and redshift distribution of the sample. We use the same resolution as the correlation function while computing $\mathbf{D}^{XY}$, i.e., $\Delta r =4~\text{\hmpc}$ spacing. However, we compute $\mathbf{D}^{XY}$ up to $r_\mathrm{max}=300~\text{\hmpc}$ as noted in the beginning of Sec.~\ref{sec:method}.
Lastly, the projection onto the Legendre multipoles is represented by another matrix operation, $\mathbf{P}^{XY}_\ell$, with $\Delta \mu=0.01$.
This operation is combined with the distortion matrix, so that our final model vector is given by $(\mathbf{P}_\ell \mathbf{D})\cdot \bm{\xi}_\mathrm{undist}$.

The undistorted model can also be represented by Legendre multipoles with the addition of an interpolation matrix in future work. The highly anisotropic nature of metal contamination makes it difficult to determine how many multipoles would be needed in this case. There are other systematics (e.g., correlated noise from data processing \cite{guyCharacterizationOfContaminants2025} and quasar redshift errors \cite{baultSystematicRedshiftErrors2025}) that we have not detailed in this section, which require similar studies.

\section{Validation with DR2 mocks\label{sec:validation}}
The DR2 \lyaf\ BAO measurement validation was built on the already extensive work for DR1 \cite{cuceuValidationDesiDr12025}. In DR2, we generated 400 synthetic realizations of the dataset using two different methods: 100 realizations with the \texttt{Saclay} method \cite{etourneauSaclayMocks2024} and 300 with the \texttt{CoLoRe-QL} method. The \texttt{CoLoRe-QL} method has improved quasar clustering on small scales and captures the BAO broadening due to non-linear structure growth, and is an improvement of the \texttt{LyaCoLoRe} method employed in DR1 \cite{farrLyaColore2020, ramirezPerezColoreSimulation2022}. These thorough validation efforts are presented in the DR2 companion paper \cite{casasValidationOfDesiDr2Lya2025}. Building on this foundation, we use only the first ten of the \texttt{CoLoRe-QL} mocks to demonstrate that the multipole representation recovers most of the information without bias.

As mentioned before, we operate only in the A region and measure the \lyaf\ auto-correlation (\lya
$\times$\lya) and the \lyaf-quasar cross-correlation (\lya$\times\mathrm{QSO}$). For our multipole compression, we consider two cases for both correlations: 1) The $\ell_\mathrm{max}=2$ case refers to using the monopole and quadrupole, and 2) the $\ell_\mathrm{max}=4$ case refers to including the hexadecapole. We find that the dipole term for \lyaxqso\ is featureless and lacks the BAO peak; therefore, we do not include it. We will discuss what dipole has to offer in Sec.~\ref{sec:discuss}, including which parameters are improved in a variation on real data. In short, the dipole does not improve the BAO constraints as expected, but provides some improvement on metal biases.
Throughout this section (and this work), we analyze the correlation function in the range $30~\text{\hmpc} < r < 180~\text{\hmpc}$ following \citetalias{desiY3LyaBAO2025}.

\begin{figure}
    \centering
    \includegraphics[width=0.97\columnwidth]{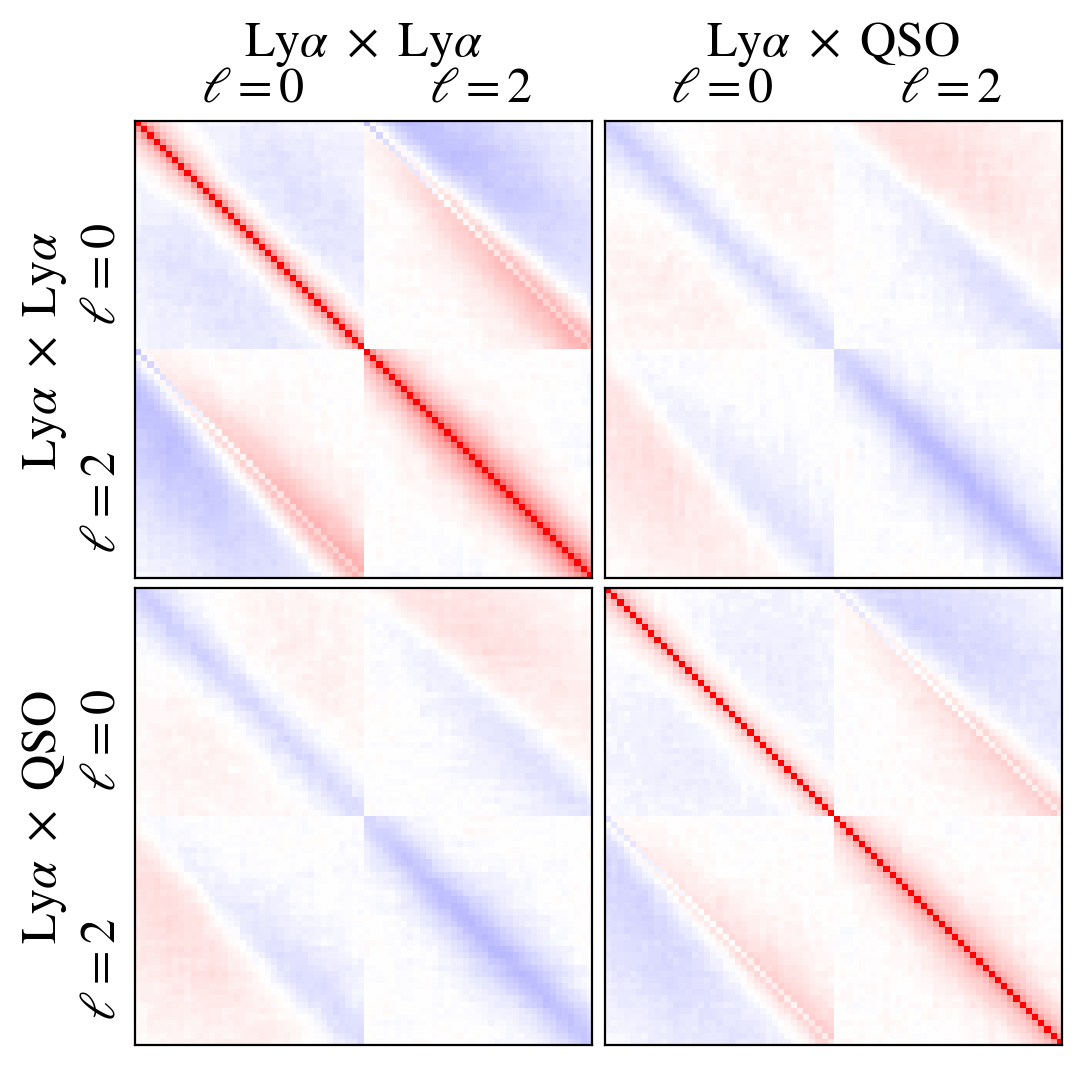} \\
    \includegraphics[width=0.97\columnwidth]{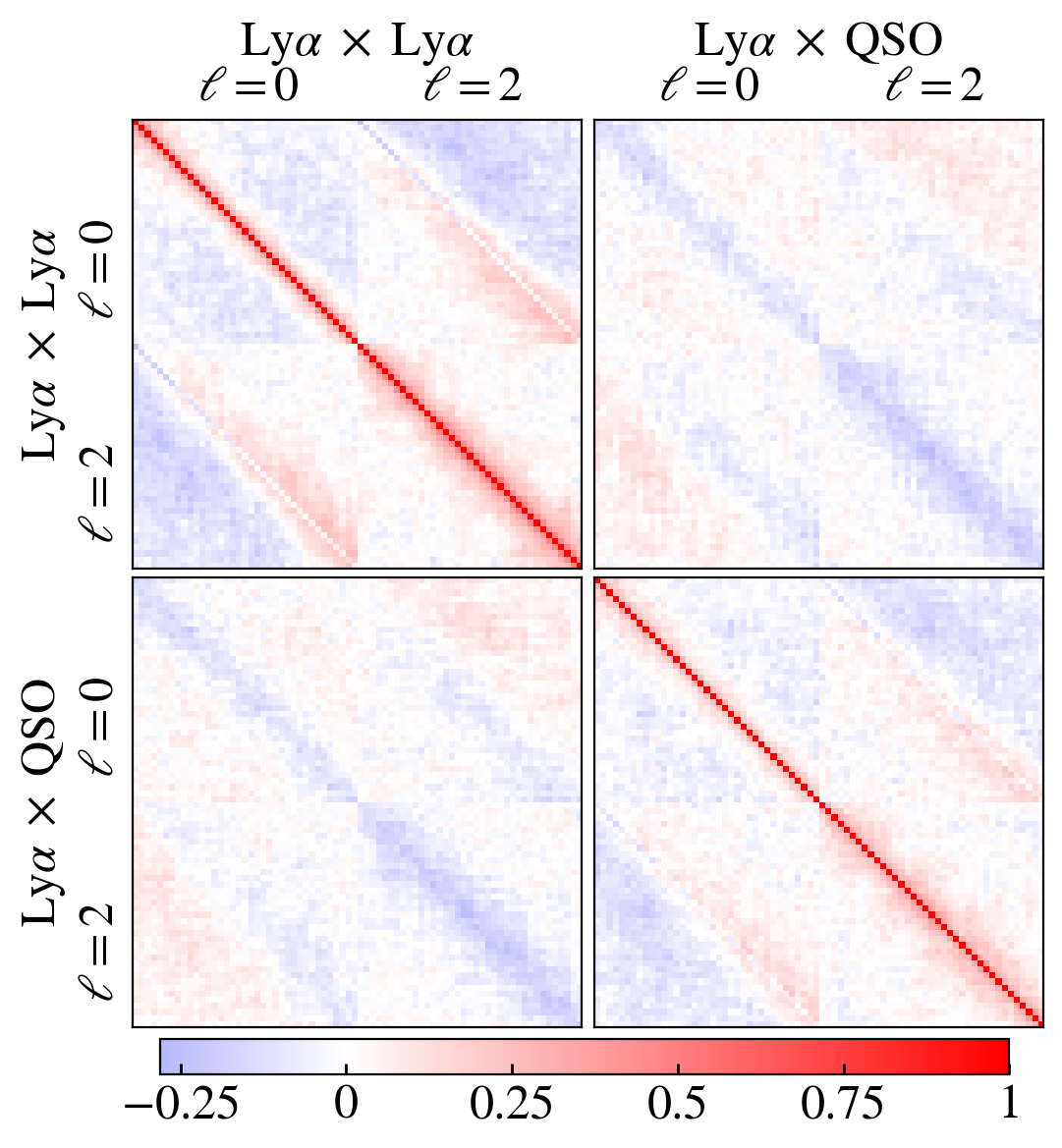}
    \caption{The average correlation matrix of ten \texttt{CoLoRe-QL} mocks ({\it top}) and the correlation matrix obtained from data ({\it bottom}). This demonstrates that mocks reasonably capture the correlations in the data covariance matrix. However, the variance in the data is considerably larger than in the mocks. Both matrices are positive definite without any ad hoc smoothing.}
    \label{fig:stack_mock_cov}
\end{figure}

\subsection{Covariance matrix}
The average of ten \texttt{CoLoRe-QL} covariances for the $\ell_\mathrm{max}=2$ case is shown in the top panel of Fig.~\ref{fig:stack_mock_cov} as a correlation matrix.
This and the $\ell_\mathrm{max}=4$ case produce a positive definite covariance matrix without any smoothing, even for one realization, as we sought. The bottom panel of this figure shows the correlation matrix obtained from actual data. We find that the median absolute difference between the two correlation matrices is $3\%$, indicating that the mocks reasonably capture the expected correlations in the data.

However, the amplitude of the diagonal is off by a large factor --- on average, the \lyaxlya\ variance is $75\%$ larger, and the \lyaxqso\ variance is $30\%$ larger in data estimates before the $15\%$ Hartlap correction (which is negligible at the $1\%$ level for the average of ten mocks). The amplitude offset is even larger for the $(r_\bot, r_\|)$ binning, by $95\%$ and $40\%$, respectively.
These offsets suggest that, if mock realizations are to be used to improve covariance matrix estimation in the future, the amplitude of the variance needs to be scaled to match the data variance. Lastly, for this section, the data covariance matrix remains positive definite, although it is noisier.

\subsection{BAO constraining power\label{sec:bao_constraining_power}}
Let us begin by assessing the potential information loss due to the projection from the full grid $(r_\bot, r_\|)$ to Legendre multipoles. We first eliminate modeling error complications and stochastic noise in measured correlations by generating a data vector at the fiducial model. In order to judge both methods on equal footing, we use covariance matrices derived from mock data, since larger statistics mitigate the non-Gaussianity of the likelihood function, and we ignore the Hartlap and Percival corrections.

\begin{figure}
    \centering
    \includegraphics[width=\columnwidth]{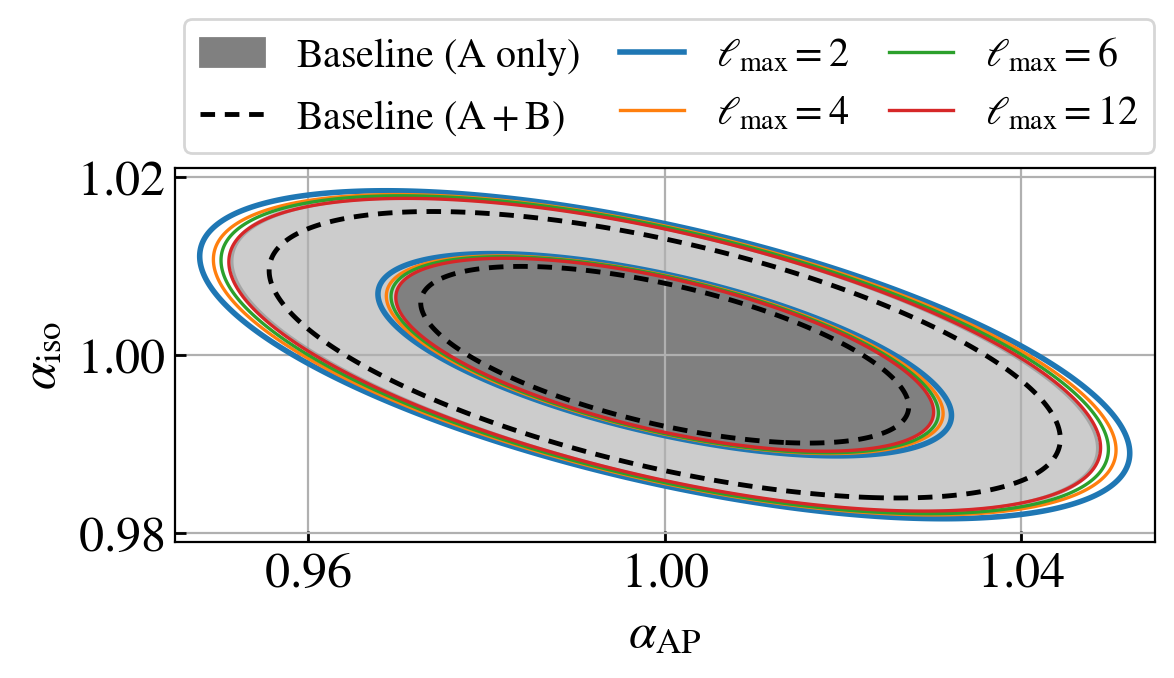}
    \caption{BAO constraining power for various analysis choices when modeling errors and noise in measured correlations are eliminated using a data vector at the fiducial model. The constraints weaken slightly when the B region is excluded ({\it gray shaded}) from the DR2 baseline analysis ({\it black dashed lines}), and weaken slightly more when using Legendre multipoles $\ell_\mathrm{max}=2$ ({\it blue solid lines}). The baseline and Legendre multipole contours converge at $\ell_\mathrm{max}=12$.}
    \label{fig:bao_forecast_mockcov}
\end{figure}

Fig.~\ref{fig:bao_forecast_mockcov} demonstrates that the information loss in the BAO parameters is small in the Legendre multipole space, and the monopole and quadrupole contain most of the BAO constraining power of the data.
The baseline DR2 analysis, which includes the B-region, constrains $\aiso$ with $0.68\%$ precision and $\aAP$ with $1.8\%$ precision. These deteriorate by $10\%$ for an A-region-only analysis. A monopole and quadrupole-only analysis (using just the A-region) does minimally worse compared to this, where $\aiso$ precision worsens by an additional $4\%$ and $\aAP$ by $7\%$. Including the hexadecapole improves both BAO constraints by only $2\%$ and $4\%$, respectively, while equivalence between the baseline and truncated multipole analyses is reached at a very high $\ell_\mathrm{max}=12$. However, the resulting data vector is still approximately seven times smaller than the baseline data vector. For completeness, the $\ell_\mathrm{max}=2$ case constrains $\aiso$ with $0.75\%$ precision and $\aAP$ with $2.1\%$ precision, and is expected to perform worse by $14.5\%$ and $17.6\%$ than the baseline DR2 choices.

\begin{figure}
    \centering
    \includegraphics[width=\linewidth]{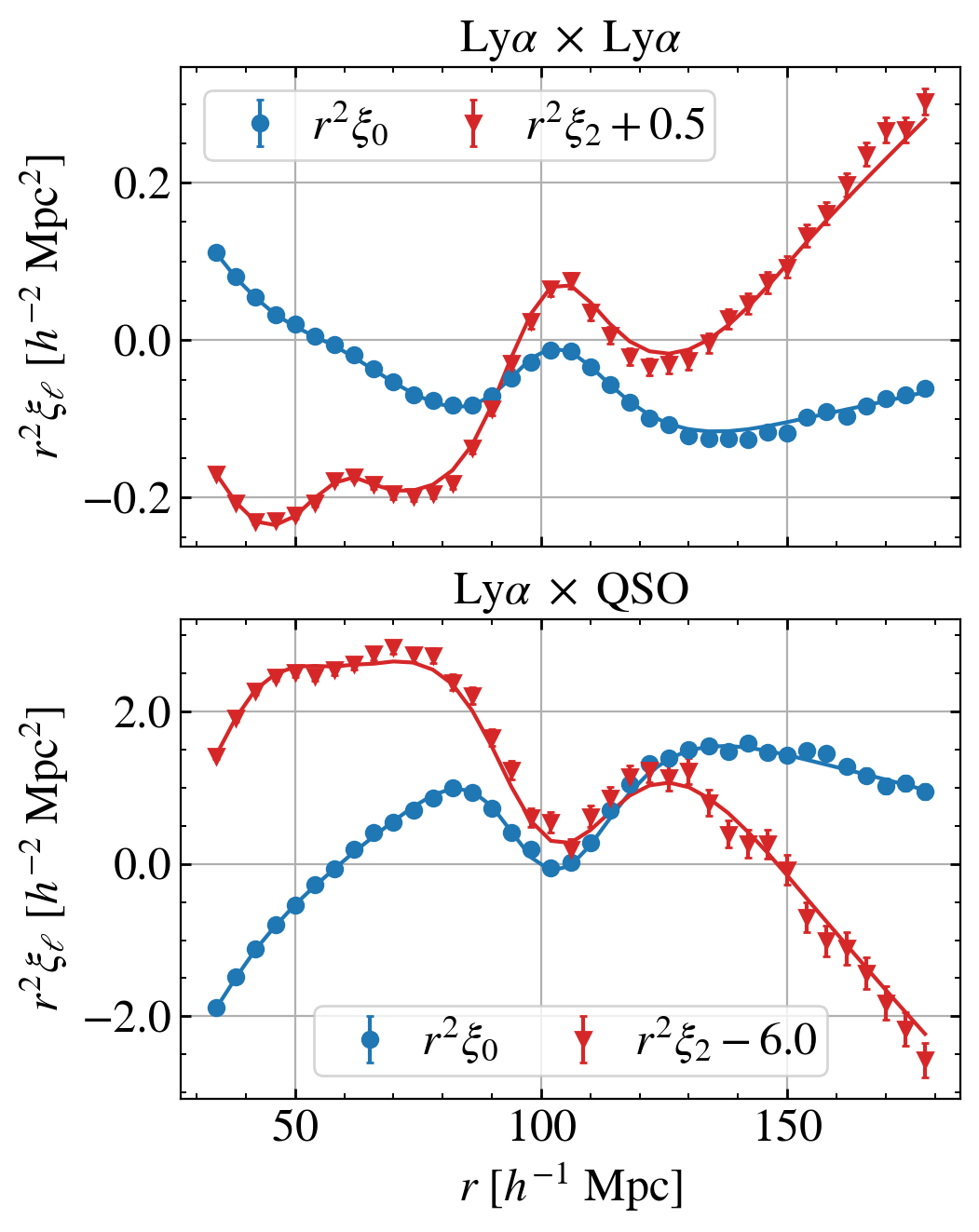}
    \caption{Average of ten \lya\ auto ({\it top}) and \lyaxqso\ cross ({\it bottom}) correlation functions measured from the mocks. The quadrupoles are shifted to match the monopole scale. The $\ell_\mathrm{max}=2$ best-fit model ({\it solid lines}) provides a good fit with PTE of $22\%$, whereas the $\ell_\mathrm{max}=4$ has a near-zero PTE.}
    \label{fig:mock_bestfit_model_vs_data}
\end{figure}

However, the constraints on all nuisance parameters become less stringent in the multipole representation with $\ell_\mathrm{max}=2$, which, in particular, underperforms at constraining the bias parameters for metal absorbers. We illustrate the correlation function with and without metal contamination in Appendix~\ref{app:metals}, which shows that the BAO peak vanishes for $\ell \ge 4$ from the correlation function, but metal peaks remain prominent. This explains how the metal contamination can be effectively isolated and quantified using higher-order multipoles. Including the hexadecapole improves these constraints by $40\%$ for the Si~\textsc{iii}~$(1207~\text{\AA})$ bias parameter and by $30\%$ for the remaining metal biases, but they are not as good as the baseline DR2 case. Furthermore, the parameter covariance matrix indicates slightly stronger correlations between BAO (most notably $\aAP$) and nuisance parameters. The correlation coefficient between $\aAP$ and the bias parameter for the  Si~\textsc{ii}~$(1260~\text{\AA})$ line rise in absolute value from $-7.4\%$ for the baseline analysis to $-15\%$ for the $\ell_\mathrm{max}=2$ case, and to $-19\%$ for the $\ell_\mathrm{max}=4$ case. This line creates a spurious peak in the correlation function at $104~$\hmpc, particularly near the BAO peak, and its correlation with $\aAP$ may pose a challenge for future analyses that use multipole decomposition.

\begin{figure}
    \centering
    \includegraphics[width=\linewidth]{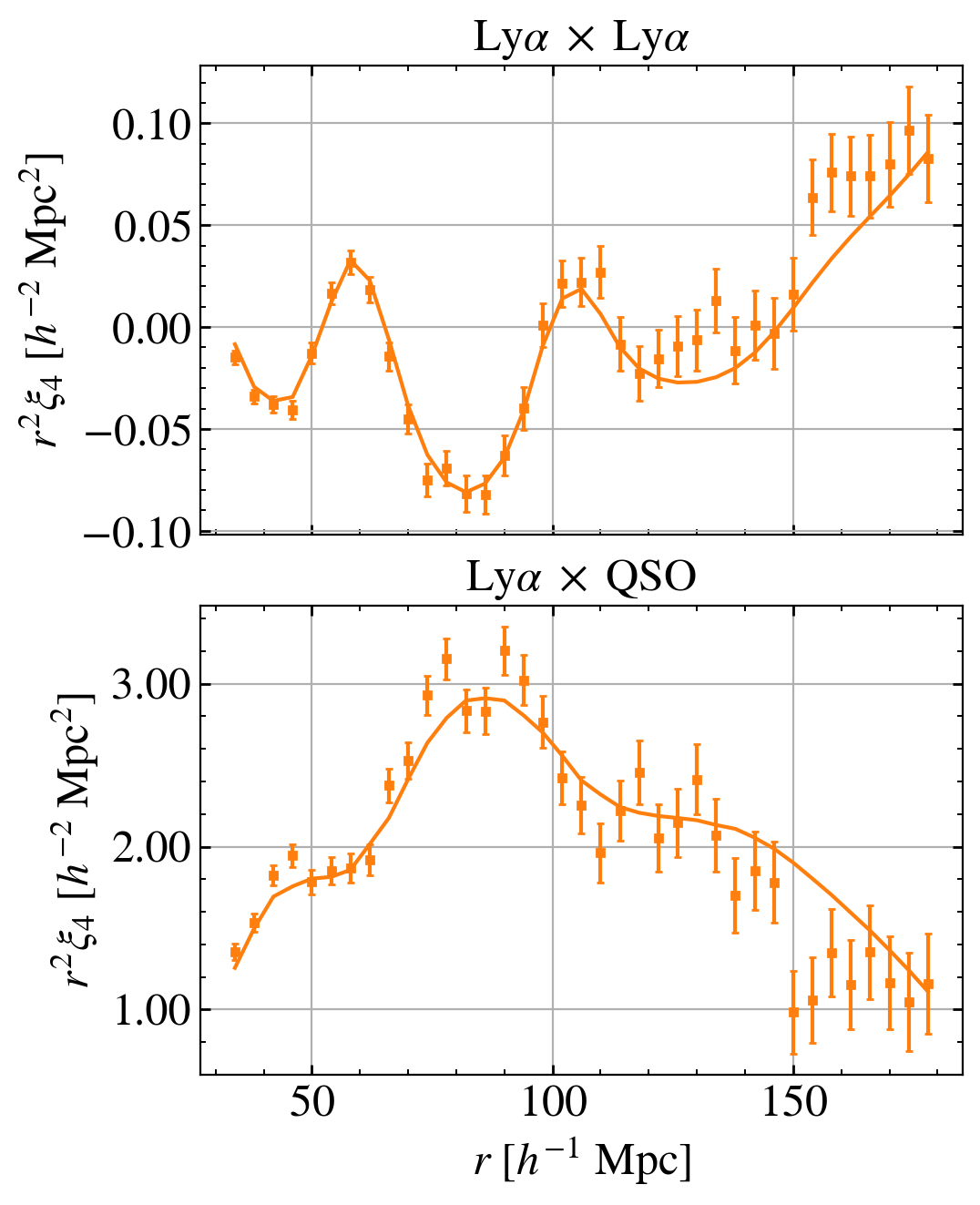}
    \caption{Average of ten \lya\ auto ({\it top}) and \lyaxqso\ cross ({\it bottom}) hexadecapoles measured from the mocks. The best-fit model ({\it solid lines}) yields a reasonable $\chi^2_\nu=36.4 / 37$ for the auto, while a poor $\chi^2_\nu=72.0 / 37$ for the cross correlation function. Including both in the fit yields a PTE of zero. Including only the auto-correlation function still performs considerably worse than the $\ell_\mathrm{max}=2$ case with a PTE of $4\%$.}
    \label{fig:mock_bestfit_model_vs_data_ell4}
\end{figure}

\begin{figure}
    \centering
    \includegraphics[width=\linewidth]{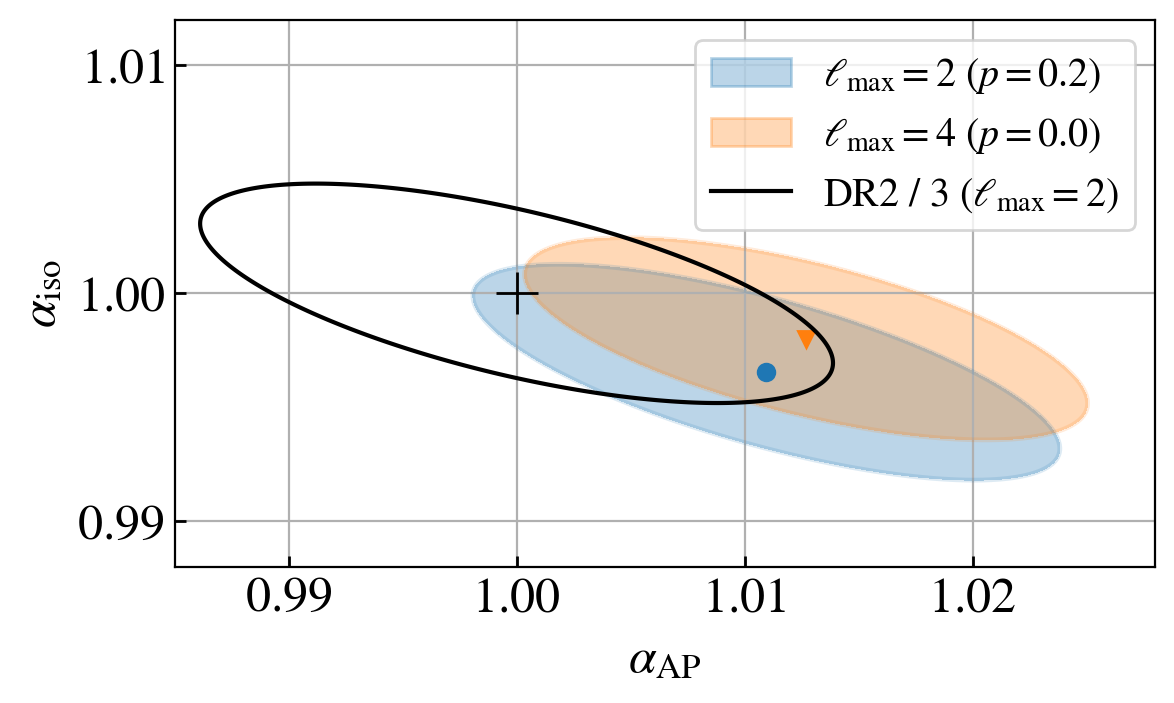}
    \caption{The best-fit BAO parameters to the average of ten correlation function measurements from the mocks. Both the $\ell_\mathrm{max}=2$ case ({\it blue}) and $\ell_\mathrm{max}=4$ case ({\it orange}) yield unbiased values, that is, one within one-third of the DR2 uncertainty using $\ell_\mathrm{max}=2$ ({\it black line}).}
    \label{fig:bao_mock_contour_valid}
\end{figure}

\subsection{Modeling accuracy}
We now turn to testing whether our modeling of the \lyaf\ multipoles is sufficiently accurate to yield an unbiased BAO measurement. We average ten correlation functions measured from the mocks to obtain a data vector that is ten times as powerful as DR2 (which would translate to $1/\sqrt{10}$ scaling in standard deviation). Similar to ref.~\cite{casasValidationOfDesiDr2Lya2025}, we set our threshold for $\aiso$ and $\aAP$ to be equal to one within one-third of the actual DR2 standard deviation, which we measure using the $\ell_\mathrm{max}=2$ setting. One-ninth of the covariance matrix derived from the minimizer establishes the contour for our validation test.

The $\ell_\mathrm{max}=2$ case provides a reasonable fit to this mock data vector with $\chi^2_\nu=148.5 / (148 - 12) = 1.092$ and a probability of having a larger value (PTE) of $22\%$. Fig.~\ref{fig:mock_bestfit_model_vs_data} shows the best-fitting model against the mock data points. \lyaxlya\ multipoles are on the top panel, and \lyaxqso\ multipoles are on the bottom panel.
Including the hexadecapole significantly degrades the goodness of fit, with $\chi^2_\nu = 287.2 / (222 - 12) = 1.37$, which has nearly zero probability ($0.03\%$) of being larger than this value. Fig.~\ref{fig:mock_bestfit_model_vs_data_ell4} displays the hexadecapole for the auto and cross correlation functions with respective best-fitting model curves in this case. This extremely poor fit is largely caused by the \lyaxqso\ hexadecapole with $\chi^2_\nu = 72 / 37 = 1.95$. Removing only this component improves the probability to $4\%$, which is still not as good as using only the monopole and quadrupole.

Finally, Fig.~\ref{fig:bao_mock_contour_valid} demonstrates that we recover unbiased BAO parameters for both cases. 
The $\alpha=1$ point remains within the one-sigma region for the $\ell_\mathrm{max}=2$ case, whereas it is marginally outside for the $\ell_\mathrm{max}=4$ case. The best-fitting BAO values are equal to one within one-third of the DR2 uncertainty for both.

We conclude this section by selecting $\ell_\mathrm{max}=2$ as our ``baseline" for the multipole analysis, since it provides a better $\chi^2$ value and an unbiased BAO measurement within one-third of the DR2 uncertainty.

\section{Results\label{sec:results}}
As our validation tests on mocks indicated, we employ the $\ell_\mathrm{max}=2$ case in our data application. In this setting, the Hartlap correction in Eq.~(\ref{eq:hartlap}) becomes 1.15 on our estimated covariance matrix, and the Percival correction in Eq.~(\ref{eq:percival}) is 1.12 on the best-fit parameter covariance. All of our reported uncertainties include this term.

\begin{figure}
    \centering
    \includegraphics[width= \linewidth]{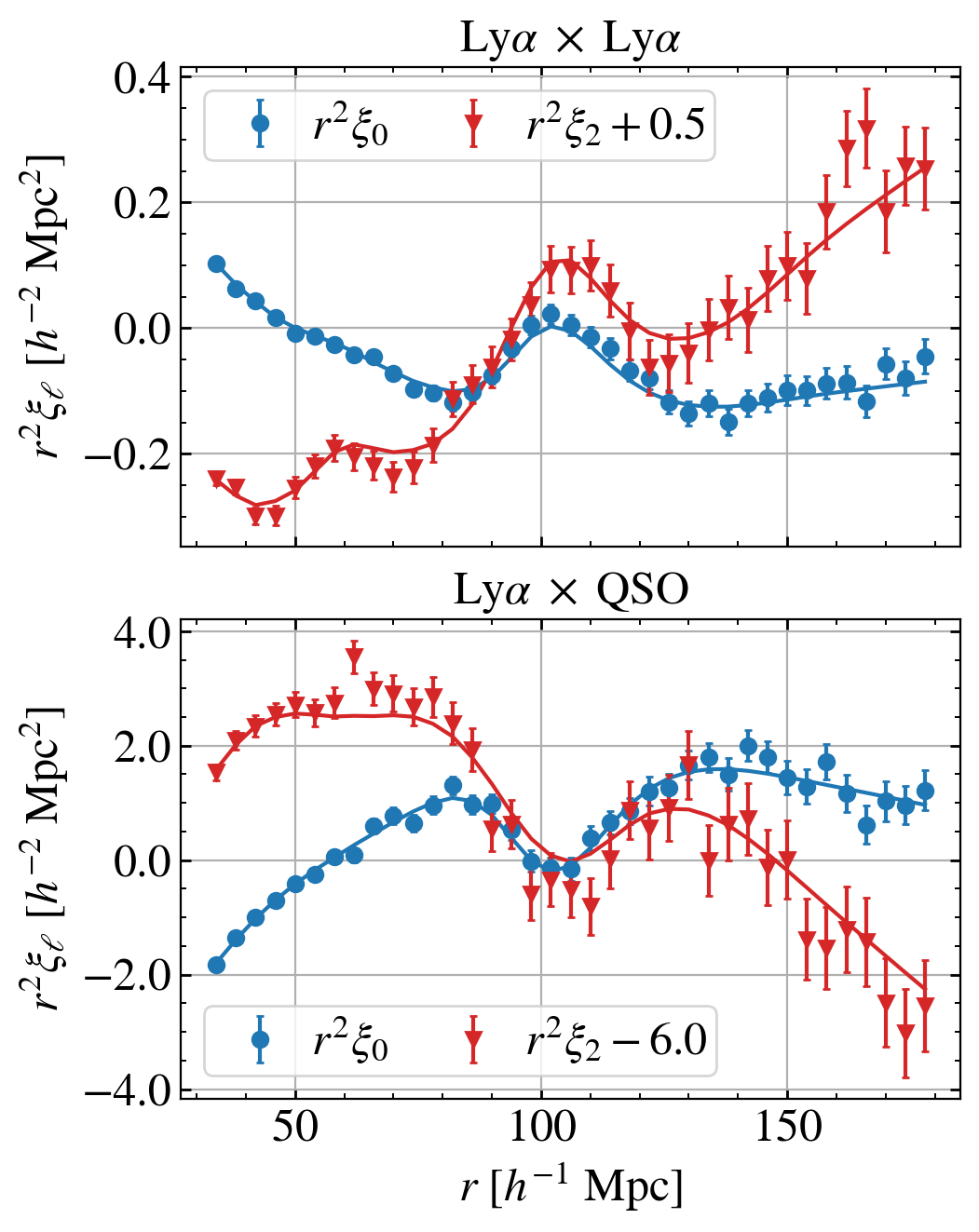}
    \caption{DESI DR2 best-fitting model vs data for \lya\ auto ({\it top}) and \lyaxqso\ cross ({\it bottom}) correlation functions. The $\ell_\mathrm{max}=2$ model provides an excellent fit with a PTE of $53\%$.}
    \label{fig:dr2_bestfit_model_vs_data}
\end{figure}

Following \citetalias{desiY3LyaBAO2025}, we adopt Gaussian priors on the effective bias of C~\textsc{iv}  systems $(b_\mathrm{C~IV(eff)})$, the quasar bias $(b_Q)$, and the systematic shift in quasar redshifts $(\Delta r_\|)$, as noted in Table~\ref{tab:all_params}. We confirmed that removing these priors does not affect the BAO measurement, as discussed in Sec.~\ref {sec:discuss}.
In addition, two nuisance parameters related to HCDs cannot be constrained in our work, unlike in \citetalias{desiY3LyaBAO2025}. Since the multipole decomposition largely weakens the constraints on the nuisance parameters, this loss of constraining power is not unexpected. These parameters are $\beta_\mathrm{HCD}$, the redshift-space distortions parameter, and $L_\mathrm{HCD}$, the width of a filter that is used to model the wings of the HCDs. As we will see shortly below, our best-fitting value for the HCD bias is zero, indicating an inability to detect the HCD contribution to the correlation function in a multipole configuration. Consequently, $\beta_\mathrm{HCD}$ and $L_\mathrm{HCD}$ become unconstrained. These parameters were constrained with the help of strong priors in previous DESI publications \cite{desiKp6BaoLya2024, desiY3LyaBAO2025} and were fixed in the eBOSS analysis \cite{bourbouxCompletedSDSSIVExtended2020}. Similarly, we fix these to the mean of the prior of \citetalias{desiY3LyaBAO2025}: $\beta_\mathrm{HCD} = 0.5$ and $L_\mathrm{HCD} = 5~$\hmpc. However, these values do not matter as their contribution is multiplied by $b_\mathrm{HCD}=0$.

\begin{figure}
    \centering
    \includegraphics[width=\linewidth]{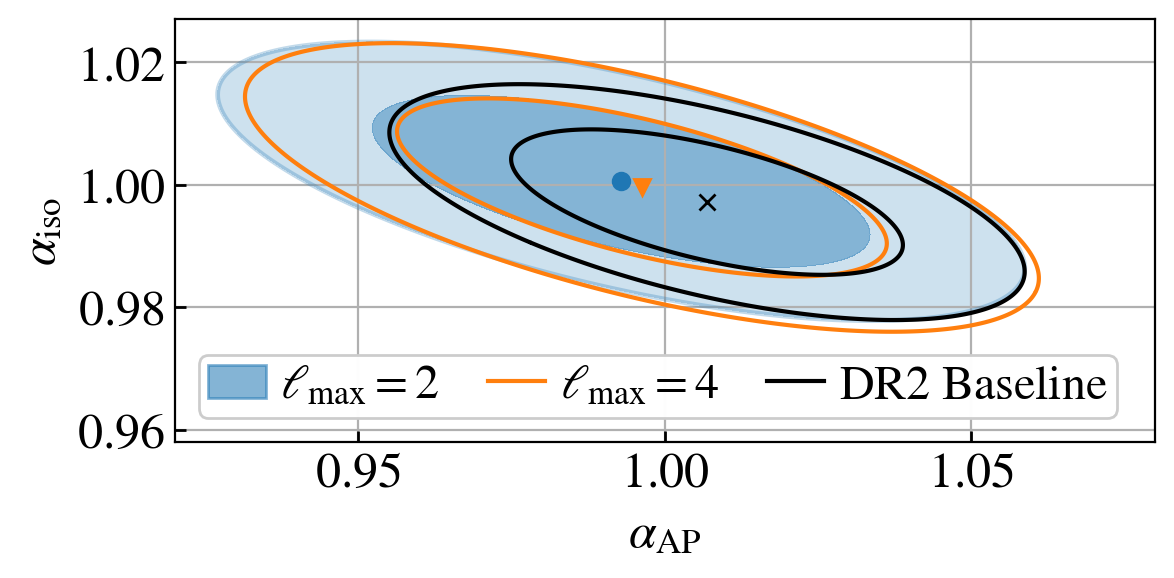}
    \caption{Our DR2 $\aiso$ and $\aAP$ measurements using $\ell_\mathrm{max} = 2$ ({\it blue}). Confidence intervals of $68\%$ and $95\%$ are plotted on equal axis scale to highlight the correlation coefficient $\rho=-0.633$. Our findings are consistent with the DESI DR2 baseline analysis ({\it black}). We perform the $\ell_\mathrm{max} = 4$ case as a variation ({\it orange}), which slightly shifts our main findings towards the DR2 baseline.}
    \label{fig:dr2_bao_contour}
\end{figure}

This leaves us with 148 data points and 15 free parameters. Our best-fitting BAO parameters are
\begin{align}
    \aiso &= 1.0005 \pm 0.0093, \\
    \aAP &= 0.9929 \pm 0.0269,
\end{align}
with a correlation coefficient $\rho=-0.623$ at an effective redshift $z_\mathrm{eff}=2.348$. We obtain an excellent fit with $\chi^2=127.7$ and a PTE of $61\%$. Fig.~\ref{fig:dr2_bestfit_model_vs_data} shows the best-fitting model against the data points for \lyaxlya\ on the top and \lyaxqso\ on the bottom panel. The fitting performance is exquisite even in the quadrupole for both correlation functions. Fig.~\ref{fig:dr2_bao_contour} shows the correlation between $\aiso$ and $\aAP$, and demonstrates the agreement between our results and the \citetalias{desiY3LyaBAO2025} results. We test the $\ell_\mathrm{max}=4$ case as a variation and find that it moves BAO parameters slightly towards \citetalias{desiY3LyaBAO2025}, which can be seen in Fig.~\ref{fig:dr2_bao_contour}. The fit in this case is worse than the $\ell_\mathrm{max}=2$ case, as expected, but unlike mock results, it has a reasonable PTE of $34\%$ with $\chi^2/\nu=214.8 / (222 - 15)$.

\begin{table}
    \caption{\label{tab:all_params}All best-fit values of our work compared to those of \citetalias{desiY3LyaBAO2025}. We quote only the statistical errors for cosmological distance ratios $D_H/r_d$ and $D_M/r_d$.}
    \begin{ruledtabular}
    \begin{tabular}{ccc}
    Parameter & This work & \citetalias{desiY3LyaBAO2025} \\
    \hline
    $\aiso$ & $1.0005 \pm 0.0093$ & $0.9971 \pm 0.0079$\\
    $\aAP$ & $0.9929 \pm 0.0269$ &$1.0069 \pm 0.0212$\\
    $\rho$ & $-0.623$ & $-0.591$ \\
    \hline
    $b_\alpha$ & $-0.180 \pm 0.008$ & $-0.135 \pm 0.007$\\
    $\beta_\alpha$ & $0.926 \pm 0.081$ & $1.445 \pm 0.064$\\
    $10^3 b_\mathrm{Si~II(1190)}$ & $-5.44 \pm 1.40$ & $-3.70 \pm 0.39$\\
    $10^3 b_\mathrm{Si~II(1193)}$ & $-0.89 \pm 1.26$ & $-3.18 \pm 0.38$\\
    $10^3 b_\mathrm{Si~III(1207)}$ & $-5.92 \pm 2.82$ & $-7.30 \pm 1.48$\\
    $10^3 b_\mathrm{Si~II(1260)}$ & $-5.33 \pm 1.70$ & $-3.67 \pm 0.40$ \\
    $10^3 b_\mathrm{C~IV(eff)}$\footnote{Prior: $\mathcal{N}(-19.0, 5.0)$} & $-18.7 \pm 5.2$ & $-18.6 \pm 4.8$ \\
    $b_\mathrm{HCD}$ & $-0.00 \pm 0.02$ & $-0.02 \pm 0.01$\\
    $b_Q$\footnote{Prior: $\mathcal{N}(3.4, 0.2)$} & $3.43 \pm 0.18$ & $3.545 \pm 0.054$\\
    $\sigma_v~[\text{\hmpc}]$ & $2.12 \pm 2.47$ & $3.18 \pm 0.63$\\
    $\Delta r_\|~[\text{\hmpc}]$\footnote{Prior: $\mathcal{N}(0.0, 1.0)$} & $0.0 \pm 1.0$ & $0.53 \pm 0.18$ \\
    $\xi_0^\mathrm{TP}$ & $0.0 \pm 2.1$ & $0.453 \pm 0.046$\\
    $10^4 a_\mathrm{noise}$ & $3.15 \pm 2.50$ & $2.21 \pm 0.15$\\
    \hline
    $z_\mathrm{eff}$ & $2.35$ & $2.33$ \\
    $D_H(z_\mathrm{eff})/r_d$ & $8.51\pm0.12$ & $8.63 \pm 0.10$\\
    $D_M(z_\mathrm{eff})/r_d$ & $39.46\pm0.65$ & $38.99 \pm 0.52$ \\
    $\rho_{HM}$ & $-0.55$ & $-0.46$\\
    \end{tabular}
    \end{ruledtabular}
\end{table}

Table~\ref{tab:all_params} lists the best-fitting values for all the nuisance parameters and compares them to the \citetalias{desiY3LyaBAO2025} results. As mentioned in previous sections, our multipole configuration provides less constraining power across the board. We have performed worse than our forecasts in Sec.~\ref{sec:bao_constraining_power} in terms of BAO precision: $\aiso$ precision has degraded by $18\%$ and $\aAP$ has degraded by $27\%$ as opposed to $14\%$ and $18\%$ respectively. Our best-fitting HCD bias ($b_\mathrm{HCD}$) is zero with 0.02 uncertainty, which is within the one-sigma region of the \citetalias{desiY3LyaBAO2025} value. We continue discussing the implications of this in Sec.~\ref{sec:discuss}. In addition to the HCD bias, the best-fitting value for the amplitude of the proximity effect ($\xi_0^\mathrm{TP}$) is also zero with a large uncertainty. Although these differences are noteworthy, they ultimately do not impact the BAO parameters, as we demonstrated in the previous section.

\section{Discussion\label{sec:discuss}}
In this work, we compressed the \lyaf\ correlation functions into Legendre multipoles and achieved our primary goal of obtaining a positive-definite covariance matrix without ad hoc smoothing. We then investigated how well these multipoles can be modeled, given the highly anisotropic nature of metal contamination and the distortions due to continuum fitting errors. We found that the monopole and quadrupole terms can be modeled well and efficiently capture BAO information, whereas including the hexadecapole yields diminishing returns, improving BAO information by only a few percent. However, nuisance parameters degrade noticeably in the multipole representation and exhibit stronger correlations between BAO parameters, particularly $\aAP$.

The Legendre compression reduces the number of data points and, therefore, the size of the covariance matrix substantially. This approach introduces an important advantage: it is now feasible to supplement covariance matrix estimation with that from mocks, as is done in galaxy clustering statistics. However, we also demonstrated that the variance of our mock data is lower than that of the real data. Consequently, the variance amplitude needs to be scaled to match the data while the mock fidelity is improved from first principles.

Our BAO parameter measurements are in complete agreement with \citetalias{desiY3LyaBAO2025}, deviating by $\delta \aiso=0.0034$ and $\delta \aAP=-0.014$, both of which are less than $\sigma/2$ differences. Our $\alpha$ measurements are slightly more correlated and have worse precision: by $18\%$ for $\aiso$ and $27\%$ for $\aAP$. The two main reasons for this are (1) not using the B region and (2) truncating the data to two multipoles for each correlation function. This is not a trivial amount of precision loss. However, the compression efficiency is noteworthy, considering we reduced the data vector by $98\%$ (from 9,306 elements to 148).

In terms of the sound horizon at the drag epoch, $r_d$, our $\alpha$ measurements correspond to the following transverse comoving distance ($D_M$), Hubble distance ($D_H= c / H$), and isotropic BAO distance ($D_V\equiv(zD_M^2D_H)^{1/3}$) measurements:
\begin{align}
    D_M(z_\mathrm{eff})/r_d &= 39.46\pm0.65, \\
    D_H(z_\mathrm{eff})/r_d &= 8.51\pm0.12, \\
    D_V(z_\mathrm{eff})/r_d &=31.46\pm0.29,
\end{align}
where $z_\mathrm{eff}=2.35$. The correlation coefficient between $D_H$ and $D_M$ is $\rho_{HM}=-0.55$. For $r_d=147.09~\text{Mpc}$, it follows that the Hubble parameter at this redshift is $H(z_\mathrm{eff})=239.5\pm3.4~\text{km~s}^{-1}~\text{Mpc}^{-1}$ and the transverse comoving distance is $D_M(z_\mathrm{eff})=5.80 \pm 0.10~\text{Gpc}$.

\begin{figure}
    \centering
    \includegraphics[width=0.9\linewidth]{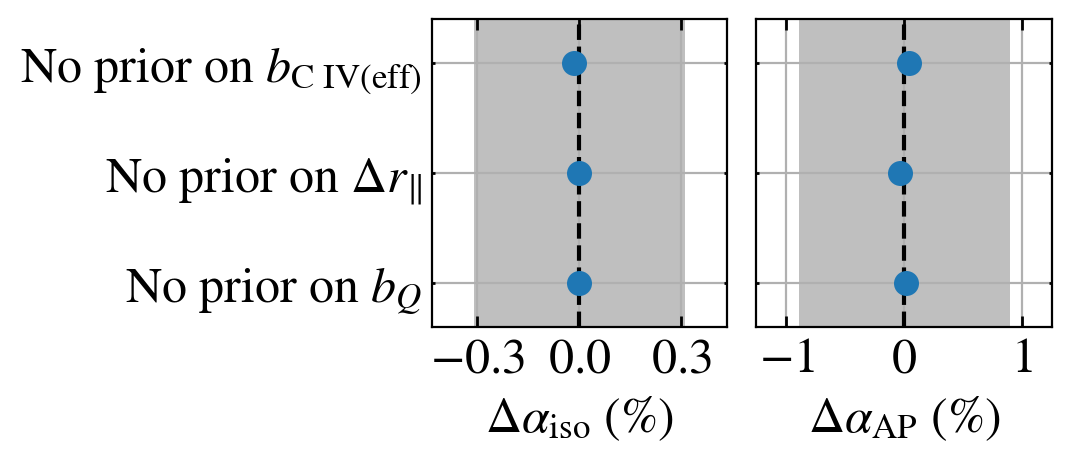}
    \caption{Shifts in the BAO parameters when a prior is removed from the analysis. The grey area represents one-third of the uncertainty in our baseline analysis. This confirms the robustness of the multipoles framework to prior choices.}
    \label{fig:prior_variations}
\end{figure}

We confirmed that our framework is robust to prior choices by removing each prior and refitting the model. The BAO measurement hardly moves, as shown in Fig.~\ref{fig:prior_variations}, where the grey shaded region corresponds to one third of the error from fits using $\ell_\mathrm{max}=2$. Our findings parallel those of \citetalias{desiY3LyaBAO2025}.

The absence of HCD bias detection in our real data analysis is accompanied by lower $\beta_\alpha$ and larger absolute $b_\alpha$ values. As we mentioned in Sec.~\ref{sec:method}, the distinguishing feature between the HCD contribution and the \lyaf\ signal is the extra smoothing along the line of sight, which becomes indistinguishable in the multipoles representation of the real data. This results in the $b_\alpha$ and $\beta_\alpha$ parameters compensating for the HCD contribution, lowering $\beta_\alpha$'s, which then raises the absolute value of $b_\alpha$.

The weaker constraints on nuisance parameters imply that analyses that exploit the full shape of the correlation functions to extract more cosmological information will require higher-order multipoles. The hexadecapole will be necessary for the full-shape analysis as it provides tighter constraints on nuisance parameters. Our forecasted improvement of $30\%$ on the metal biases from mocks remains approximately correct for the data as well. The hexacadepole improves the precision of metal biases by about $25\%$ on average. Furthermore, the precision of the quasar proximity effect improves the most, becoming $\xi_0^\mathrm{TP}=0.42 \pm 0.22$, which is an improvement of $85\%$ in precision.

When analyzing the stack of ten mocks, we found that modeling the \lyaxlya\ hexadecapole performs similarly well to the monopole and quadrupole, but modeling the \lyaxqso\ hexadecapole performs significantly worse. However, we also achieved a good fit to the data while including both hexadecapoles in the model. This can indicate two things: (1) Modeling errors are not important for the current dataset, but will gradually become important as we collect more data. (2) The model is a good description for the \lyaxqso\ hexadecapole for real data, but not a good description for simulated data. We would need to improve our simulations, focusing on quasar and metal clustering, to reliably test our model.

Relatedly, we ignored the dipole of \lyaxqso\ in this work. This term could potentially carry interesting asymmetric ``relativistic" effects such as gravitational redshift and gravitational potential terms  \cite{bonvinAsymmetricGalaxyCorFun2014, irsicRelativisticEffectsLyaForest2016, blomqvistBaoCrossLyaEboss2019}. However, in practice, it is caused by more trivial effects, such as omnipresent distortions, quasar proximity effects, systematic shift in quasar redshifts, and differences in the time evolution of quasar, forest, and metal biases. Indeed, we confirm that the \lyaxqso\ dipole is clearly detected from real data, though it is mostly featureless and cannot constrain the $\Delta r_\|$ parameter. When the dipole term is included in the analysis, metal bias constraints improve by about $10\%$ on average, while a maximum of $30\%$ improvement is achieved for $b_\mathrm{Si~III(1207)}$. However, the BAO parameters and their uncertainties remain completely unchanged. Relativistic effects can be detected using the dipole only if these trivial effects are accurately modeled.

Lastly, one might be interested in ``weighting" the multipoles in Eq.~(\ref{eq:multipole}), for example, using inverse variances of each $\xi(r, \mu)$ bin, in the hopes that it will improve the precision of best-fitting parameters by suppressing noisy $\mu$ bins. This weighting is easy to implement and propagate from model to data. However, it also breaks the orthonormality of the Legendre polynomial basis. Nevertheless, we test this formalism using $\ell_\mathrm{max}=2$ with mocks and find that it does not improve the BAO constraints over regular Legendre decomposition. In fact, performs worse for the $\aAP$ parameter. The non-orthonormality of this weighted basis also leads to increased correlations between the monopole and quadrupole.

\section*{Data Availability}
The data used in this analysis will be made public, along with DR2 (details available at \url{https://data.desi.lbl.gov/doc/releases/}). The data points shown in the figures will be available in a Zenodo repository after publication at \url{https://doi.org/10.5281/zenodo.18989744}.

\begin{acknowledgments}
This material is based upon work supported by the U.S. Department of Energy (DOE), Office of Science, Office of High-Energy Physics, under Contract No. DE–AC02–05CH11231, and by the National Energy Research Scientific Computing Center, a DOE Office of Science User Facility under the same contract. Additional support for DESI was provided by the U.S. National Science Foundation (NSF), Division of Astronomical Sciences under Contract No. AST-0950945 to the NSF’s National Optical-Infrared Astronomy Research Laboratory; the Science and Technology Facilities Council of the United Kingdom; the Gordon and Betty Moore Foundation; the Heising-Simons Foundation; the French Alternative Energies and Atomic Energy Commission (CEA); the National Council of Humanities, Science and Technology of Mexico (CONACYT); the Ministry of Science and Innovation of Spain (MICINN), and by the DESI Member Institutions: \url{https://www.desi.lbl.gov/collaborating-institutions}.

The DESI Legacy Imaging Surveys consist of three individual and complementary projects: the Dark Energy Camera Legacy Survey (DECaLS), the Beijing-Arizona Sky Survey (BASS), and the Mayall z-band Legacy Survey (MzLS). DECaLS, BASS and MzLS together include data obtained, respectively, at the Blanco telescope, Cerro Tololo Inter-American Observatory, NSF’s NOIRLab; the Bok telescope, Steward Observatory, University of Arizona; and the Mayall telescope, Kitt Peak National Observatory, NOIRLab. NOIRLab is operated by the Association of Universities for Research in Astronomy (AURA) under a cooperative agreement with the National Science Foundation. Pipeline processing and analyses of the data were supported by NOIRLab and the Lawrence Berkeley National Laboratory. Legacy Surveys also uses data products from the Near-Earth Object Wide-field Infrared Survey Explorer (NEOWISE), a project of the Jet Propulsion Laboratory/California Institute of Technology, funded by the National Aeronautics and Space Administration. Legacy Surveys was supported by: the Director, Office of Science, Office of High Energy Physics of the U.S. Department of Energy; the National Energy Research Scientific Computing Center, a DOE Office of Science User Facility; the U.S. National Science Foundation, Division of Astronomical Sciences; the National Astronomical Observatories of China, the Chinese Academy of Sciences and the Chinese National Natural Science Foundation. LBNL is managed by the Regents of the University of California under contract to the U.S. Department of Energy. The complete acknowledgments can be found at \url{https://www.legacysurvey.org/}.

Any opinions, findings, and conclusions or recommendations expressed in this material are those of the author(s) and do not necessarily reflect the views of the U. S. National Science Foundation, the U. S. Department of Energy, or any of the listed funding agencies.

The authors are honored to be permitted to conduct scientific research on Iolkam Du’ag (Kitt Peak), a mountain with particular significance to the Tohono O’odham Nation.
\end{acknowledgments}

\appendix
\section{Sample bias correction\label{app:sample_bias}}
We outline the steps to derive the sample bias correction for covariance matrix estimators belonging to the class of Eq.~(\ref{eq:cov}). That is, for a type of data that is a vector (index $i$), organized by subsamples ($s$) and associated weights ($\bm w_s$). The mean is given by $\hat {\bar x}_i = \sum_s w_{s, i} x_{s, i}/\sum_s w_{s, i}$, and the (biased) covariance estimator is given by:
\begin{equation}
    \label{eq:app_cov}\hat C^*_{ij} = \frac{\sum_s w_{s, i} w_{s, j} (x_{s, i} - \bar x_i) (x_{s, j} - \bar x_j )}{\sum_s w_{s, i} \sum_s w_{s, j}}.
\end{equation}
Without loss of generality, the weights can be normalized to one such that $\sum_s w_{s, i} = 1$ for all $i$. The simplified expressions are as follows:
\begin{align}
    \hat {\bar x}_i &= \sum_s w_{s, i} x_{s, i}, \\
    \hat C^*_{ij} &= \sum_s w_{s, i} w_{s, j} (x_{s, i} - \bar x_i) (x_{s, j} - \bar x_j ) \\
    &=  \sum_s w_{s, i} w_{s, j} (x_{s, i} x_{s, j} - \bar x_i x_{s, j} - x_{s, i} \bar x_j + \bar x_i \bar x_j).
\end{align}

We need to calculate the expectation value of this estimator $\langle \hat C^*_{ij}\rangle$ to derive its bias, assuming subsamples are statistically independent, which means the following:
\begin{equation}
    \langle x_{s, i} x_{s', j} \rangle = \begin{cases}
                    \overline{x_ix_j} & \text{if }  s = s',  \\
                     \bar{x}_i \bar{x}_j & \text{if } s \neq s',
                 \end{cases}
\end{equation}
such that the \textbf{true sample} covariance matrix is given by $C^s_{ij} = \overline{x_ix_j} - \bar{x}_i \bar{x}_j$. The covariance of the \textbf{mean} is multiplied by $\sum_s w_{s, i} w_{s, j}$ for independent and identically distributed subsamples, analgous to the variance propagation rule: $\bar x = \sum w_s x_s \Rightarrow \text{Var} (\bar x) = \sum w_s^2 \text{Var} (x_s)$. 

After calculating each term in Eq.~(\ref{eq:app_cov}), we are left with the following expression:
\begin{align}
    \langle \hat C^*_{ij} \rangle = C^s_{ij} \left[  \sum_s w_{s, i} w_{s, j} - \sum_s w^2_{s, i} w_{s, j} \right. \\
    \left.  - \sum_s w^2_{s, j} w_{s, i}  + \left( \sum_s w_{s, i} w_{s, j}\right)^2\right].
\end{align}
Note that $C^s_{ij}$ here is the covariance of subsamples, and the covariance of the mean is $C^\mathrm{mean}_{ij} = C^s_{ij} \sum_s w_{s, i} w_{s, j}$. Therefore, the bias factor of Eq.~(\ref{eq:app_cov}) is
\begin{equation}
    F_{ij} = 1 +  \sum_s w_{s, i} w_{s, j} - \frac{\sum_s w_{s, i} w_{s, j} (w_{s, i} + w_{s, j})}{\sum_s w_{s, i} w_{s, j}},
\end{equation}
such that unbiased covariance estimate is given by $\hat C^\mathrm{unbiased}_{ij} = \hat C^*_{ij} / F_{ij}$. The equal weight case ($w_s = 1/N$) recovers the sample vs population variance bias factor:
\begin{equation}
    F_{ij} = 1 + \frac{1}{N} - \frac{2/N^2}{1/N} = 1 - \frac{1}{N}.
\end{equation}
Finally, we calculate this factor and find it to be insignificant for DR2, $1-F\sim 0.0014$.

\begin{figure*}
    \centering
    \includegraphics[width=0.8\linewidth]{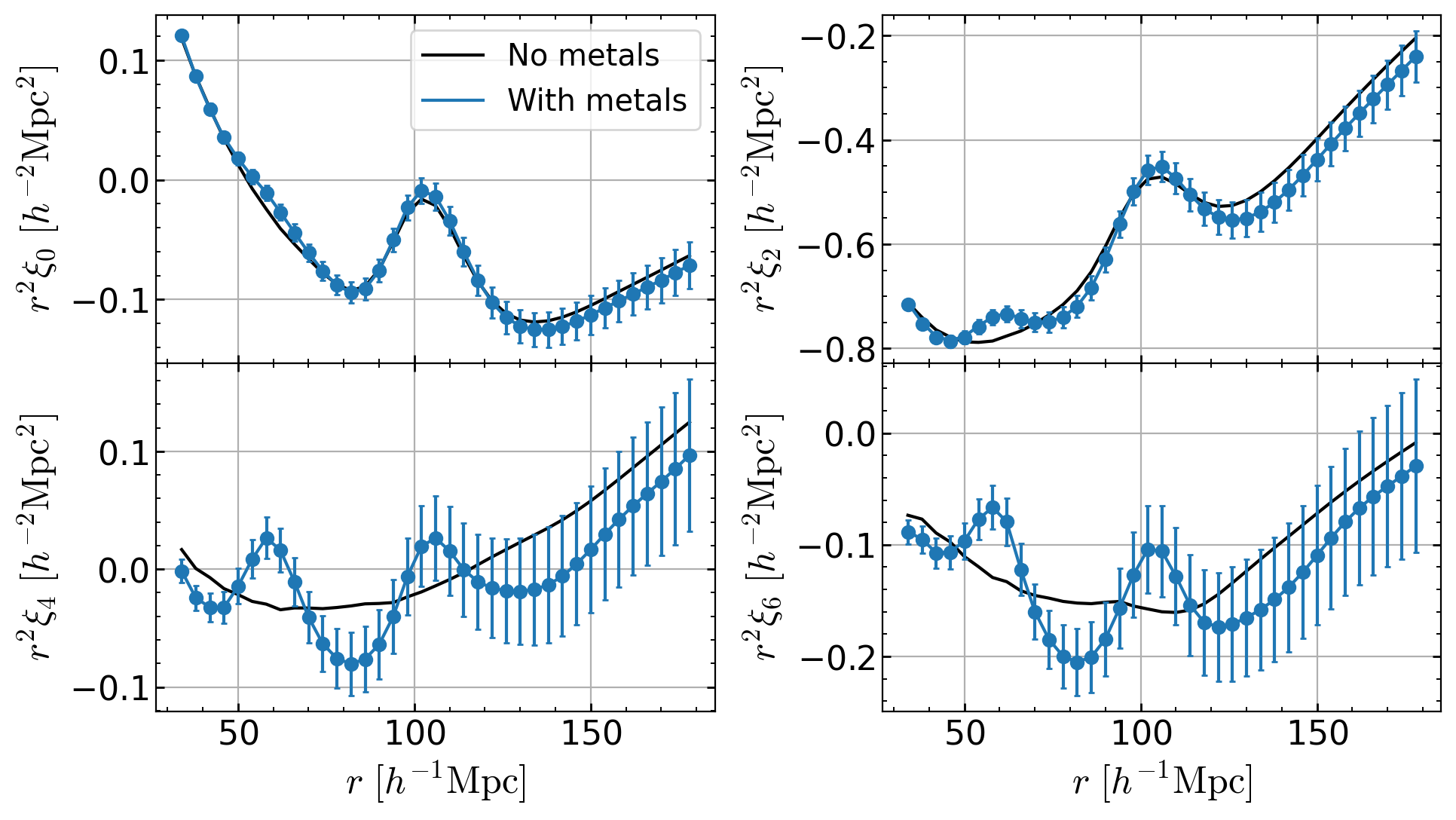}
    \caption{Multipoles of the \lyaf\ auto-correlation function for $\ell = 0, 2, 4, 6$, comparing a fiducial model without metals ({\it black curve}) and with metals ({\it blue points with error bars}) where error bars correspond to one DR2 mock realization. The BAO peak, visible at $r \approx 100\,h^{-1}\mathrm{Mpc}$ in the monopole and quadrupole, disappears for $\ell \geq 4$, while metal contamination features persist.}
    \label{fig:app_cf_auto_metals}
\end{figure*}

\section{The effect of metals in the correction function multipoles\label{app:metals}}
To illustrate the effect of metal contamination in the \lyaf\ auto-correlation function, we plot a fiducial model with and without metals. This is shown in Fig.~\ref{fig:app_cf_auto_metals}, where error bars are reflective of one mock DR2 realization. The BAO peak vanishes for $\ell \ge 4$ from the correlation function, but metal peaks remain. This is because metal contaminations are sharply localized at $\mu\approx1$, which requires higher multipoles to represent. This figure also demonstrates that since the BAO peak disappears at $\ell \ge 4$, metal contamination can be effectively isolated and quantified using higher-order multipoles.

\bibliography{references,additional_references}

\end{document}